%% file: phd.tex
\documentclass[english,twoside,12pt,a4paper,openright]{report}
\usepackage{amssymb,amsmath,mathrsfs,eufrak,babel,epsfig,fancyhdr}
\usepackage{hyperref}
\usepackage{makeidx,endnotes}
\usepackage{showidx}
\usepackage{theorem,eepic}
\usepackage{epsf,epsfig}

\def\povm#1{\mathbf{#1}}
\def\dim{\mathsf{dim}}
\def\basis{\mathbf{b}}
\def\d{\textrm{d}}

\def\vec#1{{\boldsymbol{#1}}}
\def\<{\langle}
\def\>{\rangle}
\def\kk{\rangle\!\rangle}
\def\bb{\langle\!\langle}

\def\set#1{{\sf #1}}
\def\alg#1{{\mathfrak #1}}
\def\map#1{{\mathcal{#1}}}

\def\sH{\mathscr{H}}
\def\sK{\mathscr{K}}

\def\sA{\mathscr{A}}
\def\sB{\set{B}}
\def\sS{\set{S}}
\def\sE{\set{E}}
\def\sF{\set{F}}
\def\gG{\mathbf{G}}
\def\Span{\set{Span}}
\def\Rng{\set{Rng}}
\def\Supp{\set{Supp}}
\def\Ker{\set{Ker}}

\def\Reals{\mathbb R}
\def\Cmplx{\mathbb C}

\def\dual#1{{#1}^\tau}
\def\Tr{\operatorname{Tr}}

\def\bP{{\mathbf P}}
\def\bQ{{\mathbf Q}}
\def\bM{\mathbf{M}}

\def\sT{\set{T}}
\def\sP{\set{P}}

\def\Rip{\supset_r} 
\def\Base{\mathbf{b}}
\def\Sex{S_\textrm{ex}}
\newtheorem{lemma}{Lemma}[section]

\newtheorem{definition}[lemma]{Definition}
\newtheorem{theorem}[lemma]{Theorem}
\newtheorem{remark}[lemma]{Remark}

\fancypagestyle{plain}{\fancyhead{}}

\begin{document}
\include{Titlepage}
\include{epigrafe}
\include{preface}
\tableofcontents
\include{intro}

\include{Chap1}
\include{Chap2}
\include{Chap3}
\include{Chap4}
\include{conclusions}

\include{biblio}
\end{document}

%% file: Titlepage.tex
\begin{titlepage}
\begin{center}
{\Large \bf Universit\`a degli Studi di Pavia}
\end{center}
\hrule
\begin{center}
  {\large Facolt\`a di Scienze MM.FF.NN.} \\
  \vspace{0.5cm}
  {\large Dipartimento di Fisica ``A. Volta''} \\
  \vspace{3.5cm}
  {\Huge \bf Optimization and Realization}\\
  \vspace{5mm}
  {\Huge \bf of Quantum Devices}\\
  \vspace{5cm}
  {\large PhD Thesis} \\
  \vskip 8pt
  {\large by \bf Francesco Buscemi} \\
  \vspace{2.5cm}

  \leftline{\large Supervisor: Chiar.mo~Prof. {\bf
      Giacomo~Mauro~D'Ariano}}
    
  \leftline{\large Referee: Chiar.mo~Prof. {\bf Francesco~De~Martini}}
  \vspace{1cm} {\large XVIII ciclo, A. A. 2002/2005}\\
  \hrule\vspace{0.5cm} last version: \today
\end{center}
\vfill
\eject
\end{titlepage}

%% file: epigrafe.tex
\chapter*{}
\vspace{2cm}
\begin{flushright}
\emph{Aspice convexo nutantem pondere mundum\\
terrasque tractusque maris caelumque profundum.}\vspace{.5cm}\\
Behold the world swaying her convex mass,\\
lands and spaces of sea and depth of sky.\vspace{.5cm}\\
(Vergilius, Ecloga IV\footnote{Translated from Latin by J~W~MacKail, in \emph{Virgils' Works} (Modern Library, New York, 1934).})
\end{flushright}

%% file: preface.tex
\chapter*{Preface Note}
\addcontentsline{toc}{chapter}{\bf Preface Note}

This manuscript must be intended as an informal review of the research
works carried out during three years of PhD. ``Informal'' in the sense
that technical proofs are often omitted (they can be found in the
papers) as one could do for a presentation in a public talk.  Clearly,
some background of Quantum Mechanics is needed, even if I tried to
minimize the prerequisites.

%% file: intro.tex
\chapter*{Introduction}
\addcontentsline{toc}{chapter}{\bf Introduction}

To handle information is to handle physical systems, and viceversa.
Hence, the ultimate limits in manipulating and distributing
information are posed by the very laws of physics. This is true in the
classical framework (e.~g. Landauer's principle) and in the quantum
framework, where the rules of Quantum Mechanics give rise to new---and
often not yet completely understood---restrictions and advantages to
information processing and distribution. Quantum Information Theory is
devoted to the investigation of the theoretical limits Quantum
Mechanics establishes when dealing with information encoded on quantum
systems. This thesis treats the problem of processing quantum
information\footnote{Here ``quantum information'' is a short hand for
  ``information encoded on quantum systems'' and it is basically
  equivalent to saying ``quantum states''.  Analogously, ``classical
  information'' means ``information encoded on classical systems''.}
in an optimal way by means of physically realizable devices. In fact,
linearity of Quantum Mechanics forbids basic processings of classical
information---like e.~g. copying-, broadcasting-, and
\texttt{NOT}-gates---to properly work on an unknown quantum state. The
first natural question is then: How well can we approximate such
transformations and which are the physical devices that realize these
approximations?

In contrast to its classical counterpart, quantum information is very
sensitive to noise. In a realistic setup, it is unreasonable to
completely rule out noise, since the least interaction with the
sorroundings can cause the system to be irreversibly disturbed. This
fact raises the need of designing methods to encode quantum
information in a way that is robust with respect to noise.  But in
order to do this, we have to provide a model for the noise. In this
sense, also noise can be viewed as a kind of processing of quantum
states: a ``nasty'' processing, nonetheless obeying the same laws of
Quantum Mechanics as ``good'' processings do.  The second natural
question is then: What is the role of noise in a realistic setup and
how can we control it?

In order to answer both questions, we clearly need to work in full
generality.  The appropriate mathematical tool to do this is provided
by the concept of \emph{quantum channel}. It encloses all possible
deterministic transformations of quantum states allowed by the
postulates of Quantum Mechanics. In Chapter~\ref{ch:ch1} we review the
mathematical formalism describing quantum measurements and quantum
state transformations. We face problems, such as quantum state
preparation and repeatability of quantum measurements, which, even
though they reach back to the beginnings of quantum theory, have been
revived and put into a new light by the recent developments in the
experimental techniques.

Chapter~\ref{ch:chap2} is concerned with the analysis of quantum
channels. Exploiting the convex structure of the set of channels, we
explicitly single out those that constitute the best quantum versions
of the intrisically classical copying-, broadcasting-, and
\texttt{NOT}-gates. We introduce the general theory on which such
optimization relies, and present group theoretical techniques to
analyse the common situation in which symmetries of the set of input
quantum states, after the action of the channel, propagate to the
output. This is the framework of \emph{covariant channels}. It is very
useful to describe many physical situations and, at the same time, it
permits an analytical approach.

Quantum channels are more general to describe changes of quantum
states than unitary evolutions controlled by Schr\"odinger's equation.
Nonetheless, it is well known that every quantum channel is the
transformation that a system undergoes when unitarily interacting with
an auxiliary quantum system---the so-called \emph{ancilla}---that is
discarded after the interaction took place.  Actually, this is the
only way to deterministically realize a non-unitary quantum channel.
In Chapter~\ref{ch:realization} we propose feasible implementations
for some of the channels constructed in Chapter~\ref{ch:chap2},
providing the ancillary quantum state and the global unitary
interaction. This is just a first step towards the experimental
realization which remains a far more difficult task, however, the
setup we propose to optimally copy quantum systems, in the case of
qubits (i.~e. two-levels systems) coincides with the one already used
in experiments. This is encouraging in view of a possible
generalization of experimental techniques to higher dimensional
quantum systems.

The thesis ends with Chapter~\ref{ch:noise} which deals with classical
and quantum noise. Noise is considered as acting both on the measuring
apparata and on the quantum states. More specifically, we introduce a
(partial) ordering on the convex set of measuring devices. It allows
us to characterize ``clean'' devices, namely, those which are not
affected by quantum and/or classical noise. Interestingly enough, we
show that such ordering is able to single out von Neumann's
observables as ``particularly nice'' measuring apparata. This gives an
operational characterization for the usually postulated concept of
\emph{observable}. We then focus attention on the specific model of
noise called \emph{decoherence}.  Decoherence acts destroying quantum
superpositions, thus making ineffective all quantum improvements on
the classical approach. On the other side, decoherence possesses also
foundational interest since it represents the favourite tool to
explain the quantum-to-classical transition. The process called
decoherence is actually a convex set of commuting channels satisfying
very restrictive properties. Applying techniques described in
Chapters~\ref{ch:chap2} and~\ref{ch:realization}, we provide a method
to invert decoherence and restore quantum superpositions by a feedback
control from the environment. This means that measuring a suitable
observable of the environment's degrees of freedom, and then
performing on the system a suitable unitary transformation dependent
on the measurement result, it is possible to completely cancel the
effect of decoherence.

%% file: Chap1.tex
\chapter{Quantum Measurements, Operations, and Physical
  Models}\label{ch:ch1}
\section{Classical and quantum events}
Given a finite probabilistic space $\Omega=\{1,\dots,N\}$, it is
possible to define probability distributions $P=\{p_1,\dots,p_N\}$ on
$\Omega$, where $0\le p_i\le 1$, $\sum_ip_i=1$. The set of all
probability distributions on $\Omega$, $\sP(\Omega)$, is a convex set.
It is simple to recognize its extremal points as the
delta-distributions $p_i=\delta_{ij}$. Such a structure for
$\sP(\Omega)$ can be rephrased saying that $\sP(\Omega)$ is a simplex,
namely, a convex set whose elements are \emph{uniquely} expressed as a
convex combination of extremal points. Random variables on $\Omega$
are defined as mappings $X$ from $\Omega$ into a set of ``values''
$\Upsilon$. Such values can be numbers, tensors, or whatever objects.
When $\Upsilon$ is a real vector space, it is well-defined the mean
value of $X:\Omega\to\Upsilon$, given $P\in\sP(\Omega)$, as $\bar
X\equiv \sum_ip_iX(i)$.  The set of random variables on $\Omega$ forms
a commutative algebra (under point-wise multiplication). Events are
particular random variables where $\Upsilon$ is the two-values set
$\{0,1\}$. In the classical case, \emph{events form a boolean
  algebra}\footnote{A boolean algebra $\alg B$ is a set of elements
  $\alg B=\{a,b,c,\dots\}$ satisfying the following properties: (i)
  $\alg B$ has two binary idempotent, commutative, and associative
  operations, $\land$ (logical \texttt{AND}) and $\lor$ (logical
  \texttt{OR}); (ii) $\alg B$ contains universal bounds $\varnothing$
  and $I$; (iii) for all $a\in\alg B$, there exists its complementary
  element $a'\in\alg B$ such that $a\land a'=\varnothing$, and $a\lor
  a'=I$.}: Given two events \mbox{$E_1,E_2:\Omega\to\{0,1\}$}, once
defined two binary operations $\land$ and $\lor$ as
\begin{equation}
  E_1\land E_2\equiv E_1\cdot E_2,\qquad E_1\lor E_2\equiv E_1+E_2-E_1\cdot E_2,
\end{equation}
where ``$\cdot$'' is the point-wise multiplication, it is
straightforward to verify that all properties of a boolean algebra are
satisfied.

Consider now a $N$-dimensional complex vector space $\sH$. The
analogue of probability distributions are $N\times N$ density matrices
$\rho$, i.~e. positive semi-definite trace-one matrices. The analogue
of random variables are $N\times N$ hermitian matrices $X$. Since
random variables, usually called \emph{observables}, are hermitian,
they admit a spectral decomposition $X=\sum_jx_j\Pi_j^X$, where
$\Pi_j^X$ are orthogonal projections (of rank greater than one, in
case of degeneracy). Density matrices, usually called \emph{states},
define probability distributions over the spectrum of an observable,
by means of the formula $\mu_\rho^X(x_j)\equiv\Tr[\rho\Pi_j^X]$. The
mean value of an observable $X$, given a state $\rho$, is well-defined
as $\bar X\equiv\Tr[\rho X]$. The non-commutative analogue of events
are projections $E_i=E_i^2$. The set of quantum events $\sE(\sH)$,
called \emph{quantum logic}, has two binary operations $\land$ and
$\lor$ defined as
\begin{equation}
  E_1\land E_2\equiv E_1E_2,\qquad E_1\lor E_2\equiv E_1+E_2-E_1E_2,
\end{equation}
where now the multiplication is the usual (non-commutative) matrix
multiplication.

The fundamental differences between the classical model and the
quantum model are the following (and they are basically equivalent):
\begin{enumerate}
\item the quantum logic is not a boolean algebra, since the
  distributivity law does not hold (beacause of the non-commutativity
  of the matrix product);
\item the convex set of states on $\sH$ is not a simplex, but it is
  strongly convex, whence quantum states admit many equivalent
  ensemble decompositions;
\item the algebra of observables on $\sH$ is non-commutative.
\end{enumerate}
See also the introduction paragraphs in \cite{Holevo1} and
\cite{Holevo2}.
\section{Notations}\label{subsec:notations}
To each quantum system, it is associated a complex separable Hilbert
space $\sH$, equipped with the inner product $\<\psi|\phi\>$, linear
in $\phi$ and antilinear in $\psi$, following Dirac notation. The set
of bounded operators on $\sH$ will be denoted as $\sB(\sH)$. An
operator $X$ is called \emph{self-adjoint} if it is densely defined
and $X=X^\dag$ on its domain\footnote{An operator is called
  \emph{hermitian} if its domain is dense in $\sH$ and $X\subseteq
  X^\dag$. In finite dimension the two definitions coincide and there
  is no need to bother with the density of the operator's domain.}.
Self-adjoint operators are called \emph{observables} and are in
correspondence with orthogonal resolutions of the identity by means of
the formula
\begin{equation}
X=\int_{-\infty}^{+\infty}x\d\Pi^X(x),\qquad I=\int_{-\infty}^{+\infty}\d\Pi^X(x).
\end{equation}

Positive semi-definite trace-one operator $\rho\in\sT^+(\sH)$ are
called \emph{state}. We will denote the set of states of a system
$\sH$ as $\sS(\sH)$. Since they are all compact operators, states can
be essentially viewed as infinite density matrices, also in the
infinite dimensional case, with no relevant differences from the usual
finite dimensional setting. From now on, if not otherwise specified,
we will deal with finite $d$-dimensional Hilbert spaces isomorphic to
$\Cmplx^d$, for which all linear operators are everywhere defined,
bounded and trace-class, and the self-adjointness coincides with
hermiticity.  Moreover, spectral resolutions are all discrete, i.~e.
$X=\sum_jx_j\Pi_j^X$.

Composite systems carry a tensor-product Hilbert space,
\mbox{$\sH_1\otimes \sH_2\otimes\cdots\otimes\sH_N$.} Bounded
operators $\sB(\sH)$ form themselves a Hilbert space isomorphic to
\newline\mbox{$\sH\otimes\sH\equiv\sH^{\otimes 2}$.} Once fixed a
basis $\basis=\{|i\>\}$ for $\sH$, we define the following isomorphism
between operators in $\sB(\sH)$ and vectors in $\sH^{\otimes 2}$:
\begin{equation}
  X=\sum_{ij}X_{ij}|i\>\<j|\longleftrightarrow |X\kk\equiv\sum_{ij}X_{ij}|i\>\otimes|j\>,
\end{equation}
satisfying
\begin{enumerate}
\item $\bb X|Y\kk=\Tr[X^\dag Y]$, i.~e. the Hilbert-Schmidt product;
\item $(X\otimes Y)|Z\kk= |XZY^T\kk$, where $Y^T$ denotes the
  transposition with respect to the fixed basis $\basis$;
\item $\Tr_1[|X\kk\bb Y|]=X^TY^*$, where $Y^*$ denotes the complex
  conjugation with respect to $\basis$;
\item $\Tr_2[|X\kk\bb Y|]=XY^\dag$.
\end{enumerate}
With this notation the state $|I/\sqrt{d}\kk$ is the maximally
entangled state on $\sH^{\otimes 2}$:
\begin{equation}
  \frac{1}{\sqrt d}|I\kk=\frac{1}{\sqrt d}\sum_i|i\>\otimes|i\>.
\end{equation}
Such a state will play a major role in the characterization of quantum
devices.

\section{Quantum measurements statistics: POVM's}\label{sec:POVM}
Given a state $\rho$ and an observable $X=\sum_jx_j\Pi_j^X$, the
\emph{statistical postulate} states that: The probability of obtaining
a result $x_j$ within a set $\Delta=\{x_j\}_{j\in J}$ is given by
\begin{equation}
  p(x_j\in\Delta)=\Tr\left[\rho \sum_{j\in J}\Pi_j^X\right].
\end{equation}
This means, as we already saw, that a state induces a probability
measure $\mu_\rho^X(x_j)= \Tr[\rho\Pi_j^X]$ over the set of outcomes
for a given observable.

It is clear that, apart from the actual measured value $x_j$ of the
observable $X$, the statistics of the outcomes is completely
determined by the structure of its spectral resolution $\{\Pi_j^X\}$.
In the case of an observable, such $\Pi_j$'s are orthogonal
projections, i.~e. $\Pi_i^X \Pi_j^X=\Pi_i^X\delta_{ij}$, summing up to
the identity, $\sum_j\Pi_j^X=I$.  With a little abuse of terminology,
from now on we will refer to an \emph{observable} just as a set of
orthogonal projections resolving the identity, and to the
\emph{observable outcomes} as the indices $j$'s labelling different
$x_j$'s.

The concept of observable is generalized by the concept of
\emph{positive operator-valued measure} (POVM, for short), which is a
set of positive operators $\povm P=\{P_1,P_2,\dots,P_N\}$ summing up
to the identity $\sum_iP_i=I$.  Notice that $P_i$'s need not to be
orthogonal, not even projections, and the number of oucomes of $\povm
P$, i.~e. its cardinality $|\bP|\equiv N$, can be larger than the
Hilbert space dimension $d$. As before, also in the case of POVM's,
the probability of obtaining the $j$-th outcome, given the system in
the state $\rho$, is postulated to be
$\mu_\rho^\povm{P}(j)\equiv\Tr[\rho P_j]$.

We call a two-outcomes POVM $\povm{P}=\{P,I-P\}$ an \emph{effect} or,
equivalently, a \emph{property}. According to
\cite{Busch-Lahti-Mittelstaedt} we say that an effect $\povm{P}=\{P,
I-P\}$ describes a \emph{real property} for the system $\sH$ in the
state $\rho$, if $\Tr[\rho P]=1$.

Finally, we introduce here the definition of
\emph{range}\footnote{There is no possibility of confusion between the
  range of a POVM and the range of an operator, being two completely
  unrelated concepts.} of a POVM, a concept we will extensively use in
Chapter \ref{ch:noise}.
\begin{definition}[POVM range]\label{def:POVM-range}
  Given a POVM $\povm P=\{P_1,P_2,\dots,P_N\}$, its \emph{range},
  denoted as $\Rng(\povm P)$, is defined to be the convex set of
  probability distributions $\povm p=\{p_1,p_2,\dots,p_N\}$ obtained
  as $p_i=\Tr[\rho P_i]$, varying $\rho$ in all $\sS(\sH)$.
\end{definition}
\begin{remark}\label{rem:POVM-range}
  Notice that, since $\rho$ in Definition \ref{def:POVM-range} moves
  around the whole quantum states' set, the range of a POVM identifies
  uniquely the POVM. In other words, the correspondence
\begin{equation}
\povm P\longleftrightarrow\Rng(\povm P)
\end{equation}
is one-to-one.
\end{remark}

\section{Quantum operations and
  instruments}\label{sec:state-reduction}
Since now, we dealt only with the outcomes statistics. However, in
order to completely describe the measurement statistics we need also
to specify the state reduction from prior state $\rho$ to posterior
state $\rho_j$ conditioned by the outcome $j$. The state reduction is
nothing but a rule telling us which is the system's state after the
measurement has been performed and the outcome collected.
\subsection{State collapse postulate}\label{subsec:state-collapse}
Von Neumann \cite{vonneumann} derived the well-known \emph{state
  collapse rule} starting from the following hypothesis:
\begin{enumerate}
\item the observable to be measured has discrete spectrum and it is
  non degenerate, namely all its eigenspaces are one-dimensional, in
  formula $X=\sum_ix_i|x_i\>\<x_i|$;
\item the measurement is \emph{perfectly repeatable}\footnote{See
    Section \ref{sec:repeatable}.}: Literally from von Neumann's book
  ``\emph{if a physical quantity is measured twice in succession in a
    system, then we get the same value each time}''.
\end{enumerate}
If such hypotheses are verified, then the system state after the
measurement is
\begin{equation}
\rho\longmapsto\rho_j\equiv|x_j\>\<x_j|.
\end{equation}

L\"uders \cite{luders} generalized von Neumann's theorem to degenerate
observables, introducing the postulate of \emph{minimum disturbance}
in the sense that a state, for which a property $\povm P$ is real, is
left unchanged by a measurement of $\povm{P}$. According to L\"uders'
rule, when measuring the observable $X=\sum_ix_i\Pi_i^X$ the system's
state after the measurement is
\begin{equation}\label{eq:luders}
\rho\longmapsto\rho_j\equiv\frac{\Pi_j^X\rho\Pi_j^X}{\Tr[\rho\Pi_j^X]}.
\end{equation}

The interpretation problems to which the state collapse postulate led
are beyond the aim of this manuscript.
\subsection{Quantum operations}\label{subsec:quantum-operations}
The appropriate mathematical objects describing a general quantum
state change are the so-called \emph{quantum operations} \cite{kraus}.
A quantum operation $\map E$, is a completely positive
trace-non-increasing linear mapping from $\sT^+(\sH)$ of an input
system $\sH$ to $\sT^+(\sK)$ of an output system $\sK$. The map $\map
E:\sT^+(\sH)\to\sT^+(\sK)$ is generally probabilistic, and the trace
$\Tr[\map E(\rho)]\le 1$ represents the probability that the
transformation
\begin{equation}
  \rho\longmapsto\rho'\equiv\frac{\map E(\rho)}{\Tr[\map E(\rho)]}
\end{equation}
occurs. Deterministic quantum operations, i.~e. completely positive
trace-preserving maps such that $\Tr[\map E(\rho)]=1$ for all
$\rho\in\sS(\sH)$, are called \emph{channels}. All quantum operations
admit the highly non-unique Kraus representation
\begin{equation}
\map E(\rho)=\sum_jE_j\rho E_j^\dag,
\end{equation}
where $E_j$'s are linear operators from $\sH$ to $\sK$. Nonetheless,
it is always possible to choose a Kraus representation such that
$\Tr[E_i^\dag E_j]=\|E_i\|^2_2\delta_{ij}$; we call it
\emph{canonical} Kraus representation. A quantum operation is a
channel if and only if its Kraus operators satisfy the normalization
condition
\begin{equation}\label{eq:trace-pres-with-kraus}
\sum_i E_i^\dag E_i=I.
\end{equation}
\begin{remark}\label{remark:Luders}
  The L\"uders' recipe for state change clearly corresponds to a
  quantum operation $\map E_j(\rho)=\Pi_j^X\rho\Pi_j^X$. Notice that
  the average reduced state $\bar\rho\equiv\sum_ip(i)\rho_i$ can be
  read as the output of the channel $\map
  E(\rho)=\sum_i\Pi_i^X\rho\Pi_i^X$.
\end{remark}
Every quantum operation $\map E:\sS(\sH)\to \sS(\sK)$ induces
naturally a quantum operation $\dual{\map E}$ from $\sB(\sK)$ to
$\sB(\sH)$ by means of the duality relation $\Tr[\map E(\rho)
X]=\Tr[\rho\dual{\map E}(X)]$, valid for all $\rho\in\sS(\sH)$ and
$X\in\sB(\sK)$. The map $\dual{\map E}$ is called the \emph{dual} map,
and $\dual{\map E}(I_\sK)=I_\sH$ if and only if $\map E$ is a channel.
\begin{remark}\label{remark:instrument}
  Given a measurement whose outcomes statistics is described by means
  of the POVM $\povm P$, there exist many different channels
  associated with $\povm P$. These channels are written as $\map
  E^\povm{P}(\rho)=\sum_i\map{E}_i^\povm{P}(\rho)$, with
  $\dual{(\map{E}_j^\povm{P})}(I)=P_j$, and choosing between them
  correspond to assign a particular state reduction rule.
\end{remark}
\subsection{Instruments}\label{subsec:instruments}
In the modern formulation of Quantum Mechanics, the most general tool
used to describe statistical correlations between the outcomes of
successive measurements is given by the notion of (completely
positive) \emph{instrument}, which has been introduced by Davies and
Lewis \cite{davies-lewis}. An instrument is basically a mapping
$\mathfrak{I}$ from the set $\Omega$ of outcomes to the set of quantum
operations on $\sS(\sH)$, such that $\mathfrak{I}(\bigcup_{j\in
  J}j)=\sum_{j\in J}\mathfrak{I}(j)$ and $\mathfrak{I}(\Omega)$ is a
channel. The fundamental result about instruments is the following
\cite{ozawa}
\begin{theorem}[Ozawa, 1984]
  Every statistical measurement theory, consisting both of outcomes
  statistics and state reduction rule, can be described by means of an
  appropriate instrument.
\end{theorem}
Actually, instruments formalism has been introduced in literature
mainly to handle the case of continuous outcome space $\Omega$, which
is described as a standard Borel space equipped with a
$\sigma$-algebra $\alg B(\Omega)$. When $\Omega$ is discrete and
subset of $\Reals$---as in our case---technical results become much
simpler. For further details on the general case see \cite{ozawa2}.

Finally, we define a \emph{perfect} instrument as an instrument such
that $\mathfrak{I}(j)$ is a pure contractive map, i.~e.
$\mathfrak{I}(j)(\rho)=M_j\rho M_j^\dag$, for all $j$. For example,
L\"uders instrument in Remark \ref{remark:Luders} is a perfect
instrument with $M_j=\Pi_j^X$. The instrument in Remark
\ref{remark:instrument} is perfect only if $\map E_j^\povm
P(\rho)=\sqrt{P_j}\rho\sqrt{P_j}$.

\section{Physical realizations}\label{sec:physical-real}
Intruments provide both outcomes statistics and state reduction due to
a measurement process. Implicitly, we assume that such a measurement
is nondestructive, in the sense that the system is left in a state
conditioned by the outcome and not, for example, absorbed by a counter
or a calorimeter. The only reasonable way to look for an
implementation of a nondestructive measuring process on a quantum
system $\sH$ is to engineer an indirect measurement scheme. This means
that we make the system interact with an apparatus $\mathscr{A}$ and,
after some time, we measure an observable $Y$ on the apparatus. In
formula:
\begin{equation}\label{eq:indirect-meas}
  \mathfrak{I}(j)(\rho)=\Tr_\mathscr{A}\left[\left(I_\sH\otimes\Pi_j^Y\right)U(\rho\otimes|a\>\<a|)U^\dag\right],
\end{equation}
where $\{\Pi_j^Y\}_j$ is an orthogonal resolution of $I_\mathscr A$
coming from the diagonalization of $Y\in\sB(\mathscr A)$. Clearly such
a procedure gives rise to an instrument, as described in the previous
Section.  Ozawa \cite{ozawa} proved the converse:
\begin{theorem}[Ozawa, Indirect Measurement, 1984]
  Every instrument $\mathfrak{I}$ admits an indirect measurement scheme
  as in Eq. (\ref{eq:indirect-meas}).
\end{theorem}
The correspondence is not one-to-one: there are many
different---though statistically equivalent---indirect measurement
schemes producing the same instrument; conversely, given the indirect
measurement scheme, the resulting instrument is unique.
\subsection{Levels of description of quantum measurements}
There are basically three ways to describe the statistical aspects of
quantum measurements, depending on the level of details required:
\begin{enumerate}
\item One is interested only in the outcome statistics. Then the
  maximum generality lies in the concept of POVM, as we saw in Section
  \ref{sec:POVM}. Notice that, given the outcome statistics for all
  quantum states, the POVM is defined uniquely---see Remark
  \ref{rem:POVM-range}.
\item Also the state reduction rule is requested. The notion of
  instrument encloses all possible cases, see Section
  \ref{sec:state-reduction}. Evidently, many different instruments
  produce the same outcome statistics, i.~e. they all correspond to
  the same POVM.
\item The most detailed description characterizes even the state of
  the apparatus, the physical interaction between the system and the
  apparatus, and the observable to be measured on the apparatus.
  Clearly the same instrument is obtainable by means of different
  indirect measurement schemes.
\end{enumerate}
Summarizing, given the outcome statistics, the POVM is uniquely
defined. Given the POVM, there are many instruments describing it.
Similarly, there are many indirect measurement schemes realizing a
given instrument. The choice between different equivalent physical
realizations of a measurement process can be made according only to
``practical'' considerations.
\subsection{Example: standard coupling}
Consider a discrete observable $X=\sum_ix_i\Pi_i^X$ of the system
$\sH$ in initial state $|\psi\>$ and let the apparatus system be
$\sA=L^2(\Reals)$ in initial state $|\phi_a\>$. Now, let $\sH$ and
$\sA$ interact in such a way that the the observable $X$ couples with
the apparatus' momentum $P_a$. This means that the unitary operator is
\begin{equation}
  U=e^{-i\lambda X\otimes P_a}.
\end{equation}
The momentum operator $P_a$ is the generator of translations, in the
sense that
\begin{equation}
  e^{-\frac i\hslash x_0P_a}\phi_a(x)=\phi_a(x-x_0),
\end{equation}
where $\phi_a(x)=\<x|\phi_a\>$. Hence, the initial system+apparatus
state $|\psi\>\otimes|\phi_a\>$ evolves as
\begin{equation}
  U|\psi\>\otimes\phi_a(x)=\sum_i\Pi^X_i|\psi\>\otimes\phi_a(x-\hslash\lambda x_i),
\end{equation}
and, by making assumptions on the value of the coupling constant
$\lambda$ and the initial state $|\phi_a\>$, it is always possible to
obtain functions $\phi_a(x-\hslash\lambda x_i)$ with (almost) disjoint
supports. In other words, it is always possible to model an
interaction between system and apparatus such that the indirect
measurement is a position measurement on the apparatus---the usual
``pointer position'' measurement.
\subsection{Example: embedding and optimal phase measurement}
From the geometrical point of view, the unitary interaction of the
system with a fixed ancilla state in the indirect measurement scheme
(\ref{eq:indirect-meas}) simply corresponds to a linear (isometrical)
embedding of the system $\sH$ into a composite Hilbert space
$\sH\otimes\sA$.  The measurement on the apparatus then defines a
conditional expectation from $\sH\otimes\sA$ to $\sH$, giving rise to
probability and state reduction. An embedding into a larger space can
always be described by means of an isometry $V$, i.~e. a bounded
operator such that $V^\dag V=I$. If the input system state is $\rho$,
then the embedded state---that is, the system+apparatus state
\emph{after} the interaction---is $U(\rho\otimes|a\>\<a|)U^\dag\equiv
V\rho V^\dag$.

In Ref.~\cite{idphase}, we exploited an embedding for single-mode
states of the electromagnetic field in order to achieve a physical
realization of the optimal phase measurement. It is well known that
the phase of the electromagnetic field does not correspond to any
self-adjoint operator. Quantum estimation theory \cite{Holevo1,
  helstrom} provides the optimal POVM for the phase measurement in
terms of Susskind-Glogower operators
\begin{equation}\label{eq:susskind}
\d\hat\mu(\phi)=\frac{\d\phi}{2\pi}|e^{i\phi}\>\<e^{i\phi}|,\quad\int_0^{2\pi}\d\hat\mu(\phi)=I,
\end{equation}
where $|e^{i\phi}\>\equiv\sum_{n=0}^\infty e^{i\phi\hat n}|n\>$. The
optimal phase measurement outcomes distribution is then
\begin{equation}
  \d\mu_\rho^\phi=\frac{\d\phi}{2\pi}\<e^{i\phi}|\rho|e^{i\phi}\>.
\end{equation}

Using the double-ket notation introduced in
Section~\ref{subsec:notations}, consider now the eigenstates of the
hetherodyne photocurrent $Z=a-b^\dag$
\begin{equation}
\hat Z|D(z)\kk=z|D(z)\kk,
\end{equation}
where $D(z)=e^{za^\dag-z^*a}$ are the displacement operators,
satisfying the completness relation
\begin{equation}
  \int_\mathbb C\frac{\d^2z}{\pi}|D(z)\kk\bb D(z)|=I^{\otimes 2}.
\end{equation}
The following isometry
\begin{equation}
V=\frac{1}{\sqrt{2\pi}}\int_\mathbb C\d^2\alpha f(|\alpha|)|D(\alpha)\kk\<e^{i\arg\alpha}|,\quad\int_0^\infty\d t|f(t)|^2=\frac 1\pi
\end{equation}
embeds a single-mode state into a two-modes state in such a way that,
measuring the hetherodyne photocurrent
\begin{equation}
\begin{split}
  p(z)&=\frac 1\pi\Tr\left[V\rho V^\dag\ |D(z)\kk\bb D(z)|\right]\\
  &=\frac 12\left|f(|z|)\right|^2\<e^{i\arg z}|\rho|e^{i\arg z}\>,
\end{split}
\end{equation}
one obtains the optimal phase distribution $\d\mu_\rho^\phi$ as the
marginal of $p(z)$ on the variable $\phi=\arg z$. Notice that here we
are performing a joint measurement on both modes, not just an indirect
measurement on the second mode. However, the form of the embedding $V$
provides a natural way to implement the phase POVM
(\ref{eq:susskind}).
\section{Repeatable measurements}\label{sec:repeatable}
In Subsection \ref{subsec:state-collapse} we introduced the von
Neumann-L\"uders state collapse principle, derived from the hypothesis
of discreteness of spectrum, repeatability, and minimum disturbance.
In what follows, we derive all the consequences that arise from the
only hypothesis of repeatability, thus obtaining the most general form
of a repeatable measurement. See \cite{repmeas} for a detailed
derivation.

First of all, why should we focus on repeatable measurements? Clearly,
there are a lot of natural measurement schemes which are far from
being repeatable, think of e.~g. a photon counter or a fluorescent
screen at the end of a Stern-Gerlach apparatus. In the past decades,
however, technology of quantum experiments improved in such a way that
nondestructive measurements on individual atomic objects are quite a
common task, see e.~g. one atom micro-masers and ions traps.

In the modern formulation of Quantum Mechanics, repeatability
hypothesis has lost the \emph{in-principle} relevance it enjoyed in
the early foundational books as von Neumann's. Nowadays, repeatability
is understood just as a property which characterizes some particular
measurement processes. More precisely, repeatable measurements are
related to \emph{preparation} procedures. In fact, preparing a quantum
system in a particular state means preparing it in a state having some
pre-specified \emph{real property}, as defined in Section
\ref{sec:POVM}. For example, in order to prepare the pure state
$|\psi\>$, one may take a collection of quantum systems and perform
over them a repeatable measurement of the effect described by the POVM
$\{|\psi\>\<\psi|,I-|\psi\>\<\psi|\}$. Of course, the preparation
succeeds when the outcome $|\psi\>\<\psi|$ comes out. Then, the von
Neumann-L\"uders state collapse rule tells us that the state of the
system after measurement is in fact the pure state $|\psi\>$. In this
sense, repeatable measurements have often been regarded as
measurements of observables---projective orthogonal resolution of the
identity---causing a collapse of the state on one of their
eigenvectors.

In \cite{repmeas} we showed that there exist repeatable measurements
which give rise to nonorthogonal POVM's and, moreover, which \emph{do
  not even admit any} eigenvector, that is to say, the reduced state
is different at every repetition of the measurement.  This result
makes a clear separation between the concepts of repeatability,
preparation and reality in Quantum Measurement Theory.

The starting point is the hypothesis of repeatability. A first
consequence of this is due to Ozawa \cite{ozawa}:
\begin{theorem}[Ozawa, Repeatable Measurements, 1984]
  An instrument satisfies repeatability hypothesis only if it has
  discrete spectrum.
\end{theorem}
Then, perfect\footnote{For the definition of perfect instruments, see
  Subsection \ref{subsec:instruments}.} repeatable instruments are
described by a set of contractions $\{M_j\}$ such that $\sum_iM_i^\dag
M_i=I$ and
\begin{equation}
\frac{\|M_jM_k|\psi\>\|}{\|M_k|\psi\>\|}=p(j|k)=\delta_{jk},
\end{equation}
for all $j,k$ and all $|\psi\>\in\sH$. The only technical point
recalled in the paper is that, allowing for infinite dimensional
Hilbert spaces, one has also to deal with properties of operators such
as closeness. In our case, all $M_i$'s are bounded and everywhere
defined, hence closed \cite{halmos}. A close operator possesses closed
range and kernel (the support is always closed since by definition it
is the orthogonal complement of the kernel) and the Hilbert space
$\sH$ can hence be decomposed as
\begin{equation}
\sH\simeq\Ker(M_j)\oplus\Supp(M_j)\simeq\Rng(M_j)\oplus\Rng(M_j)^\perp,\qquad\forall j.
\end{equation}
The consequences are the following:
\begin{enumerate}
\item All ranges, for different outcomes, must be orthogonal, i.~e.
  \begin{equation}
\Rng(M_i)\perp\Rng(M_j),\quad i\neq j.
\end{equation}
\item All ranges must be contained in respective supports, i.~e.
  \begin{equation}
\Rng(M_i)\subseteq\Supp(M_i),\quad\forall i.
\end{equation}
\item All $M_i$'s satisfy the condition
  \begin{equation}
M_i^\dag M_i|_{\Rng(M_i)}\equiv I_{\Rng(M_i)}.
\end{equation}
\end{enumerate}
Now, the fundamental difference between operators on finite and
infinite Hilbert spaces is that, in finite dimension,
$\Supp(X)\simeq\Rng(X)$ always, while, in the infinite dimensional
case, one can have $\Supp(X)\subset\Rng(X)$ or, viceversa,
$\Rng(X)\subset\Supp(X)$, strictly. This holds basically because in
the infinite dimensional case there exist proper subspaces with the
same dimension as the whole Hilbert space $\sH$. This observation lead
us to the following:
\begin{theorem}
  For finite dimensional systems, only observables admit repeatable
  measurement schemes, and the system state collapses according to the
  von Neumann-L\"uders rule (\ref{eq:luders}).
\end{theorem}
So the finite dimensional case describes precisely what one usually
expects about the structure of repeatable measurements. It is
nonetheless possible to construct a simple example in infinite
dimension, enclosing all counter-intuitive features of the infinite
dimensional case.  Let us consider a two-outcomes POVM:
\begin{equation}\label{eq:repeatable-povm}
\begin{split}
&P_0=p|0\>\<0|+\sum_{j=0}^\infty|2j+1\>\<2j+1|,\\
&P_1=(1-p)|0\>\<0|+\sum_{j=0}^\infty|2j+2\>\<2j+2|.
\end{split}
\end{equation}
Notice that $\povm P=\{P_0,P_1\}$ is a nonorthogonal measurement. We
can describe such a POVM by means of the following instrument:
\begin{equation}\label{eq:repeatable-scheme}
\begin{split}
&M_0=\sqrt{p}|1\>\<0|+\sum_{j=0}^\infty|2(j+1)+1\>\<2j+1|,\qquad M_0^\dag M_0=P_0,\\
&M_1=\sqrt{1-p}|2\>\<0|+\sum_{j=0}^\infty|2(j+1)+2\>\<2j+2|,\qquad M_1^\dag M_1=P_1,
\end{split}
\end{equation}
in the sense that, got the $i$-th outcome, the state changes as
$M_i\rho M_i^\dag$. Repeatability hypothesis can be simply checked.
Analysing the structure of scheme (\ref{eq:repeatable-scheme}) one
recognizes a unilateral-shift behaviour of the kind $S|n\>=|n+1\>$.
Actually, this unilateral-shift structure is a general feature of
nonorthogonal repeatable measurements. Since $S$ does not admit any
eigenvector, analogously the scheme (\ref{eq:repeatable-scheme})
changes the system state at every repetition of the measurement and
there are no states which are left untouched by such a scheme. In
other words, in infinite dimensional systems there exist repeatable
measurements which \emph{cannot} satisfy minimum disturbance
hypothesis, even in principle, and hence \emph{cannot} be viewed as
preparation procedures.


%% file: Chap2.tex
\chapter{Characterization and Optimization of Quantum
  Devices}\label{ch:chap2}
In order to handle information encoded on quantum states we need to
engineer astonishingly precise and accurate devices since the least
loss of control in manipulating quantum systems can lead to extremely
detrimental effects on the whole process. The theoretical
investigation is the starting ground in designing such optimal quantum
devices. This Chapter is devoted, first, to giving a complete and
tractable characterization of quantum channels, second, to exploiting
such characterization to single out optimal devices according to
particular figures of merit that we will introduce and explain from
time to time.

The basic assumption we will adopt is to consider input quantum states
belonging to sets obeying some symmetry constraints---i.~e.
satisfying invariance properties under the action of some groups of
transformations. Moreover, we will choose figures of merit conforming
in a natural way to the same symmetry constraints.  These two
conditions lead to the very well established mathematical framework of
\emph{covariant channels}, for which the characterization simplifies,
making explicit calculations analytically solvable. Actually,
covariant channels form convex sets whose structure is (in some cases)
known and optimal devices lie on the border of such sets. In this way,
the problem resorts to a semi-definite linear program.

In particular, we will focus on channels optimally approximating the
impossible tasks of copying, broadcasting, and performing \texttt{NOT}
on unknown quantum states. The symmetries we will deal with are
universal symmetry (invariance under the action of $\mathbb{SU}(d)$),
phase-rotations symmetry (invariance under the action of
$\mathbb{U}(1)^{\times d}$), and invariance under the group of
permutations.

\section{Choi-Jamio\l kowski isomorphism}\label{sec:choi-jam}
A useful tool to characterize quantum channels in finite dimensional
systems is the Choi-Jamio\l kowski \cite{jamiol,choi,nonuniv}
isomorphism---one-to-one correspondence---between channels \mbox{$\map
  E:\sS(\sH)\to\sS(\sK)$} and positive operators $R_\map E$ on
$\sK\otimes\sH$ defined as follows:
\begin{equation}\label{eq:choi-jam}
R_\map E=(\map E\otimes \map I)|I\kk\bb I|\longleftrightarrow\map E=\Tr_\sH\left[\left(I\otimes\rho^T\right)R_\map E\right],
\end{equation}
where $\map I$ is the identity map on $\sS(\sH)$,
$|I\kk=\sum_i|i\>\otimes|i\>$ is the maximally entangled (non
normalized) vector in $\sH\otimes\sH$, and $O^T$ denotes the
transposition with respect to the fixed basis used to write $|I\kk$.
Different Kraus representations for $\map E(\rho)=\sum_iE_i\rho
E_i^\dag$ correspond to different ensemble representations for $R_\map
E=\sum_i|E_i\kk\bb E_i|$, the canonical\footnote{See Subsection
  \ref{subsec:quantum-operations}.} being the diagonalizing one.
Trace-preservation constraint $\sum_iE_i^\dag E_i=I_\sH$ rewrites as
$\Tr_\sK\left[R_\map E\right]=I_\sH$.

Choi-Jamio\l kowski isomorphism (\ref{eq:choi-jam}) turns out to be
very useful in describing covariant channels. In the following Section
we shall recall some basic notions about group theory.
\section{Group-theoretical techniques}\label{sec:group-techniqes}
\subsection{Elements of group theory}
A unitary (projective) representation on $\sH$ of the group $\povm G$
is a homomorphism $\povm G\ni g\mapsto U_g\in\sB(\sH)$, with $U_g$
unitary operator, such that the composition law is preserved:
\begin{equation}
U_gU_h=\omega(g,h)U_{gh}.
\end{equation}
The cocycle $\omega(g,h)$ is a phase, i.~e. $|\omega(g,h)|=1$, for all
$g,h\in\povm G$, and it satisfies the relations
\begin{equation}
\begin{split}
&\omega(gh,k)\omega(g,h)=\omega(g,hk)\omega(h,k)\\
&\omega(g,g^{-1})=1.
\end{split}
\end{equation}
A unitary representation is called \emph{irreducible} (UIR) if there
are no proper subspaces of $\sH$ left invariant by the action of all
its elements. Two irreducible representations $U^1$ and $U^2$ of
$\povm G$ on $\sH_1$ and $\sH_2$, respectively, are called
\emph{equivalent} if there exists a unitary $T:\sH_1\to\sH_2$ such
that $TU^1_g=U^2_gT$, for all $g\in\povm G$. The fundamental result
concerning UIR's of a group is the following:
\begin{lemma}[Schur]\label{lem:schur}
  Let $U^1$ and $U^2$ be two UIR of $\povm G$ on $\sH_1$ and $\sH_2$,
  respectively. Let $B:\sH_1\to\sH_2$ a (bounded) operator such that:
\begin{equation}
BU^1_g=U^2_gB,
\end{equation}
for all $g\in\povm G$. Then:
\begin{enumerate}
\item $U^1$ and $U^2$ equivalent $\Longrightarrow$ $B\propto T$;
\item $U^1$ and $U^2$ inequivalent $\Longrightarrow$ $B=0$.
\end{enumerate}
\end{lemma}
\begin{remark}[Abelian groups]
  From Schur Lemma simply follows that fact that, if the group $\povm
  G$ is abelian, nemely $g_1g_2=g_2g_1$ for all $g_1,g_2\in\povm G$,
  then all its UIR's are one-dimensional. In fact, all $U_g$'s must be
  proportional to the same unitary operator $T$ and they are all
  simultaneously diagonalizable, hence reducible on direct sums of
  one-dimensional invariant subspaces.
\end{remark}
\subsection{Invariant operators and covariant channels}
Let $W_g$ a reducible unitary representation of $\povm G$ on $\sH$.
Then $\sH$ can be decomposed into a direct sum of minimal invariant
subspaces:
\begin{equation}\label{eq:direct-sum_decomposition}
\sH\simeq\bigoplus_i\sH_i.
\end{equation}
Each $\sH_i$ supports one UIR of $\povm G$. Some UIR's can be
equivalent or inequivalent. Let us group equivalent UIR's under an
index $\mu$ labelling different equivalence classes, and let an
additional index $i_\mu$ span UIR's among the same $\mu$-th
equivalence class. Since equivalent UIR are supported by isomorphic
subspaces, i.~e. $\sH_{i_\mu}\simeq\sH_{j_\mu}\simeq\sH_\mu$ for all
$i_\mu,j_\mu$ in the same $\mu$-th class, we can rewrite the
decomposition (\ref{eq:direct-sum_decomposition}) as
\begin{equation}\label{eq:wedderburn}
  \sH\simeq\bigoplus_\mu\sH_\mu\otimes\Cmplx^{d_\mu},
\end{equation}
where $d_\mu$ is the cardinality (degeneracy) of the $\mu$-th
equivalence class.  Decomposition (\ref{eq:wedderburn}) is usually
called \emph{Wedderburn's decomposition} \cite{zhelobenko}, the spaces
$\sH_\mu$ are called \emph{representation spaces}, and the spaces
$\Cmplx^{d_\mu}$ \emph{multiplicity spaces}. Then, the following
decomposition for the representation $W_g$ holds
\begin{equation}\label{eq:wedderburn2}
  W_g=\bigoplus_\mu W_g^\mu\otimes I_{d_\mu}.
\end{equation}
With Eq.~(\ref{eq:wedderburn2}) at hand, it is simple to derive the
form of an operator $B$, invariant under the action of the reducible
representation $W_g$, i.~e.
\begin{equation}
  W_g^\dag BW_g=B,\qquad\forall g\in\povm G.
\end{equation}
Since the above implies $[B,W_g]=0$, for all $g$, then:
\begin{equation}\label{eq:invariant-form}
  B=\bigoplus_\mu I^\mu\otimes B_{d_\mu},
\end{equation}
where $B_{d_\mu}$ is an operator on $\Cmplx^{d_\mu}$. In other words,
the operator $B$ is in a block-form since it cannot connect
inequivalent representations and can act non-trivially only on
multiplicity spaces of the representation $W_g$. This is precisely
what is contained in the Schur's Lemma \ref{lem:schur}.

Now, consider a family of quantum states $\sF\subseteq\sS(\sH)$ that
is invariant\footnote{Notice that this requirement is weaker than
  requiring that $\sF$ is the orbit of a single seed state under the
  action of $\mathbf{G}$.} under the action of a group $\mathbf{G}$,
namely $U_g\rho U_g^\dag\in\sF$ for all $g\in\mathbf{G}$ and all
$\rho\in\sF$. The group, and then the family $\sF$, can be discrete as
well as continuous. A channel $\map E:\sF\to\sS(\sK)$ is said to be
covariant under the action of the group $\povm G$ if
\begin{equation} 
\map E(U_g\rho U_g^\dag)=V_g\map E(\rho) V_g^\dag,\qquad\forall g\in\povm G,
\end{equation}
where $U_g$ and $V_g$ are two generally reducible unitary
representations of $\povm G$ on $\sH$ and $\sK$, respectively. In a
sense, the channel $\map E$ is ``transparent'' with repect to the
action of the group $\povm G$ and the image of the invariant family
$\sF$ is another invariant family $\map E(\sF)$. Using Choi-Jamio\l
kowski isomorphism (\ref{eq:choi-jam}), the above covariance condition
for $\map E$ rewrites as an invariance condition for $R_\map E$
\cite{nonuniv}, namely,
\begin{equation}
[R_\map E,V_g\otimes U_g^*]=0,\qquad\forall g\in\povm G,
\label{eq:invariance}\end{equation}
where, as usual, the complex conjugate is with respect to the basis
used to write $R_\map E$. Decomposing $\sK\otimes\sH=\bigoplus_\mu
\sH_\mu \otimes \Cmplx^{d_\mu}$, one gets:
\begin{equation}
  R_\map E=\bigoplus_\mu I^\mu\otimes R_{d_\mu},
\end{equation}
with positive blocks $R_{d_\mu}$.

Another direct consequence of Eq. (\ref{eq:invariant-form}) is the
form of a group-averaged operator, namely
\begin{equation}
\<X\>_\povm G\equiv\int_\povm G\d gU_gXU_g^\dag,\qquad\int_\povm G\d g=1.
\end{equation}
Clearly, $\<X\>_\povm G$ is invariant, whence, if the representation
$U_g$ of $\povm G$ decomposes the Hilbert space as
$\sH=\bigoplus_\mu\sH_\mu\otimes\mathbb{C}^{d_\mu}$, it can be written
as
\begin{equation}\label{eq:group-average}
\<X\>_\povm G=\bigoplus_\mu I^\mu\otimes\frac{\Tr_{\sH_\mu}[X]}{\dim\sH_\mu}.
\end{equation}
Notice that $\Tr_{\sH_\mu}[X]$ is a short-hand notation for
$\Tr_{\sH_\mu}[P_\mu XP_\mu]$, where $P_\mu$ is the projection of
$\sH$ onto $\sH_\mu\otimes\mathbb{C}^{d_\mu}$.
\subsection{Example:
  $\mathbb{SU}(d)$-covariance}\label{subsec:sud-covariance}
A typical $\mathbb{SU}(d)$-covariance, also known as \emph{universal}
covariance, for short U-covariance, is that under the representation
of many input and output copies, namely when
$\sH\equiv(\mathbb{C}^d)^{\otimes N}$ and
$\sK\equiv(\mathbb{C}^d)^{\otimes M}$, with $U_g\equiv W_g^{\otimes
  N}$ and $V_g\equiv W_g^{\otimes M}$. Here, $W_g$ is the defining
representation of $\mathbb{SU}(d)$, and invariance condition
(\ref{eq:invariance}) reads:
\begin{equation}\label{eq:SUd-invariance}
\left[R_\map E,W_g^{\otimes M}\otimes(W_g^*)^{\otimes N}\right]=0.
\end{equation}

The general Wedderburn's decomposition for such a representation is
very complicated and channels satisfying covariance
(\ref{eq:SUd-invariance}) will be studied with a somewhat different
approach, see Subsection \ref{subsec:universal-cloning}. Nonetheless,
there are two situations in which universal covariance can be
conveniently faced using $R_\map E$ machinery. The first situation is
when $U_g\equiv W_g$ and $V_g\equiv W_g^*$. This is the case in which
we are requiring a \emph{controvariance} condition:
\begin{equation}
\map E(W_g\rho W_g^\dag)=W_g^*\map E(\rho)W_g^T.
\end{equation}
The invariance condition reads $\left[R_\map E,W_g^{\otimes
    2}\right]=0$, which implies $R_\map E=r_SP_S^{(2)}+r_AP_A^{(2)}$,
where $P_S^{(2)}$ and $P_A^{(2)}$ are respectively the projections
onto the totally symmetric and the totally antisymmetric subspaces of
$\sH^{\otimes 2}$. We will analyze this case in Subsection
\ref{subsec:UNOT}.

The second situation is when $d=2$, namely when we deal with qubits.
First of all, in this case the two representations $W_g$ and $W_g^*$
are equivalent, since \mbox{$W_g^*=\sigma_yW_g\sigma_y$} \cite{jones}.
Hence the Wedderburn's decomposition for $W_g^{\otimes
  M}\otimes(W_g^*)^{\otimes N}$ is the same as for
$W_g^{\otimes(M+N)}$ which the well-known Clebsch-Gordan series
\cite{messiah} for the defining representation of
$\mathbb{SU}(2)$\footnote{Rigorously speaking, this is not the
  Wedderburn's decomposition since different $\sH_J$ can support
  \emph{equivalent} representations. See Subsection
  \ref{subsec:Usuperbro}.}:
\begin{equation}\label{eq:Clebsch-Gordan-series}
\begin{split}
  (\mathbb{C}^2)^{\otimes M}\otimes(\mathbb{C}^2)^{\otimes M}&\simeq 
\underbrace{\bigoplus_{j=j_0}^{M/2}\bigoplus_{l=l_0}^{N/2}}_{\simeq\bigoplus_\mu}\underbrace{\left(\mathbb{C}^{2j+1}\otimes\mathbb{C}^{2l+1}\right)}_{\simeq\sH_\mu}\otimes\underbrace{\left(\mathbb{C}^{d_j}\otimes\mathbb{C}^{d_l}\right)}_{\simeq\mathbb{C}^{d_\mu}}\\
&\simeq\bigoplus_{j=j_0}^{M/2}\bigoplus_{l=l_0}^{N/2}\bigoplus_{J=|j-l|}^{j+l}\sH_J\otimes\mathbb{C}^{d_j}\otimes\mathbb{C}^{d_l},
\end{split}
\end{equation}
where $j_0,l_0$ are equal to 0 or 1/2 if $M,N$ are even or odd,
respectively, and
\begin{equation}\label{eq:CB-multiplicities}
  d_j=\frac{2j+1}{M/2+j+1}\binom{M}{M/2-j}.
\end{equation}
We will analyze this case in Subsection \ref{subsec:Usuperbro}.
\subsection{Example: $\mathbb{U}(1)$-covariance}
The defining representation of $\mathbb{U}(1)$ is simply a phase
$e^{i\phi}\in\mathbb C$. In higher dimensions, we can impose either
\emph{phase-covariance} \cite{Holevo1}, that is,
\begin{equation}
U_\phi\equiv e^{i\phi N},\qquad N=n|n\>\<n|,\quad n=0,\dots,d-1,
\end{equation}
useful to model systems driven by a Hamiltonian with equally spaced
energy levels, as the harmonic oscillator Hamiltonian, or
\emph{multi-phase covariance}, that is, covariance under a unitary
representation of the $d$-fold direct product group
$\mathbb{U}(1)\times \dots\times\mathbb{U}(1)$:
\begin{equation}\label{eq:multi-phase-rotation}
U_\vec\phi\equiv\sum_{n=0}^{d-1}e^{i\phi_n}|n\>\<n|,\qquad\phi_n\in[0,2\pi[,
\end{equation}
where $\vec\phi=\{\phi_n\}$ is a vector of $d$ \emph{independent}
phases. Notice that one of such phases is actually an overall phase
and can be disregarded: for a $d$-dimensional system we then have
$(d-1)$ effective phase-degree of freedom. In the following we shall
adopt multi-phase covariance, and, where there is no possibility of
confusion, we shall interchange the terms phase-covariance and
multi-phase covariance.  Notice that, in the case of qubits, the two
concepts coincide.

Also phase-covariance is typically applied to many copies of input and
output (say $N$ and $M$, respectively). For qubits the representation
$U_\phi^{\otimes N}$ decomposes as (see Eq.~(\ref{eq:wedderburn2}))
\begin{equation}
  U_\phi^{\otimes N}\simeq\bigoplus_{l=l_0}^{N/2}e^{i\phi J_z^{(l)}}\otimes I_{d_l},
\end{equation}
where $J_z^{(l)}=\sum_{n=-l}^ln|l,n\>\<l,n|$ is the angular momentum
component along rotation axis, say $z$-axis, in the $l$
representation.  As in the universal case---$\mathbb{U}(1)$ is a
subgroup of $\mathbb{SU}(2)$, actually---dealing with two-dimensional
systems allows us to handle the complete Wedderburn's decomposition
and work in full generality, even with mixed sates (see
Subsection~\ref{subsec:phasebroad}).

In higher dimensional systems, we shall restrict ourselves to pure
input states. This implies that the many-copies input state lives
actually in the totally symmetric subspace\footnote{This is true only
  for many-copies pure input states $\psi^{\otimes N}$. Indeed, a
  many-copies mixed input state $\rho^{\otimes N}$ is generally non
  symmetric.} $\sH\equiv(\mathbb{C}^d)^{\otimes N}_S$. Moreover,
optimal map will be found to have output supported in
$\sK\equiv(\mathbb{C}^d)^{\otimes M}_S$. Now, a convenient way to
decompose the composite space $\sK\otimes\sH$ in the Wedderburn's form
$\bigoplus_\mu\sH_\mu\otimes\mathbb{C}^{d_\mu}$, is the following:
\begin{equation}
\sK\otimes\sH\simeq\bigoplus_{\{m_i\}}\sH_{\{m_i\}}\otimes\sH,
\end{equation}
where $\{m_i\}$ is a multi-index such that $\sum_im_i=M-N$. Invariant
subspaces are clearly one-dimensional, since the group is abelian, and
equivalence classes are spanned by\footnote{We consider here only
  maximally degenerate equivalence classes, namely, equivalence
  classes whose degeneracy equals the dimension of the input Hilbert
  space $(\mathbb{C}^d)^{\otimes N}_S$. For example, the vector
  $|1\>^{\otimes M}\otimes|0\>^{\otimes N}$ supports an irrep but it
  cannot be written as in Eq. (\ref{eq:span-equiv-class-phcov}). In
  Subsection \ref{subsec:phasecloning} we will see how this constraint
  indeed does not cause a loss of generality.}:
\begin{equation}\label{eq:span-equiv-class-phcov}
  \sH_{\{m_i\}}\otimes\sH=\Span\Big\{|\{m_i+n_i\}\>\otimes|\{n_i\}\>\Big\}_{\{n_i\}}.
\end{equation}
In the above equation, $\{n_i\}$ is a multi-index such that
$\sum_in_i=N$. The vectors $|\{n_i\}\>$ are defined as:
\begin{equation}\label{eq:symmetric-vectors}
|\{n_i\}\>=\frac{1}{\sqrt{N!}}\sum_\tau\Pi_\tau^N|\underbrace{0,\dots,0}_{n_0},\dots,\underbrace{d-1,\dots,d-1}_{n_{d-1}}\>\Pi_\tau^N,
\end{equation}
where $\Pi_\tau^N$ are permutations of the $N$ systems. In other
words, $|\{n_i\}\>$ are totally symmetric normalized states, whose
occupation numbers are denoted by the multi-index $\{n_i\}$. Clearly,
by varying $\{n_i\}$ over all possible values $0\le n_i\le N$, the set
$|\{n_i\}\>$ spans all input space $\sH$. Analogous arguments hold for
the vectors $|\{m_i+n_i\}\>$ in $\sK$. That the decomposition using
$|\{m_i+n_i\}\>\otimes|\{n_i\}\>$ is useful to identify the block
structure of a multi-phase covariant channel is clear noticing that
\begin{equation}
  U_\vec\phi^{\otimes M}\otimes(U_\vec\phi^*)^{\otimes N}|\{m_i+n_i\}\>\otimes|\{n_i\}\>=e^{i\sum_im_i\phi_i}|\{m_i+n_i\}\>\otimes|\{n_i\}\>,
\end{equation}
for all possible choice of $\{n_i\}$. We'll make use of this
decomposition in Subsections \ref{subsec:phasecloning} and
\ref{subsec:phasenot}.
\subsection{Example: permutation group invariance}
Most channels of physical interest act on input states which are
indeed ``many-identical-copies states''. This is the case, for
example, of estimation channels, which optimally reconstruct an
unknown input state by performing measurements on $N$ copies of it.
Analogously, when the task is distributing quantum information to $M$
users, typically one requires that the reduced state is the same for
each user. Both situations can be described by saying that input
and/or output states are actually permutation invariant states. In
formula:
\begin{equation}
\map E(\rho)=\map E(\Pi_\tau^N\rho\Pi_\tau^N)=\Pi_\sigma^M\map E(\rho)\Pi_\sigma^M,\qquad\forall\tau,\sigma,
\end{equation}
where $\Pi_\tau^N$ and $\Pi_\sigma^M$ are
(real\footnote{Representations of permutations are always real.})
representations of the input and output spaces permutations,
respectively. When both properties are satisfied, the operator $R_\map
E$ must equivalently satisfies the following invariance condition:
\begin{equation}
[R_\map E,\Pi_\sigma^M\otimes\Pi_\tau^N]=0.
\label{eq:perm-inv-0}\end{equation}
Notice that such an invariance property is stronger than that in
Eq.~(\ref{eq:invariance}) since it implies both conditions
\begin{equation}
  \left[R_\map E,\Pi_\sigma^M\otimes I^{\otimes N}\right]=0\qquad\left[R_\map E,I^{\otimes M}\otimes\Pi_\tau^N\right]=0,
\end{equation}
for all $\sigma,\tau$, hence in particular for $\sigma=\tau$. The
fundamental tool that comes now at hand is the so-called
\emph{Schur-Weyl duality} between permutation group representations on
qubits and the defining representation of $\mathbb{SU}(2)$. The
duality relation tells that $\Pi_\sigma^M$ decomposes
$(\mathbb{C}^2)^{\otimes M}$ precisely as $W_g^{\otimes M}$, namely,
\begin{equation}
  (\mathbb{C}^2)^{\otimes M}\simeq\bigoplus_{j=j_0}^{M/2}\mathbb{C}^{2j+1}\otimes\mathbb{C}^{d_j},
\end{equation}
but with exchanged role for the spaces. Explicitly,
$\mathbb{C}^{2j+1}$ is now the multiplicity space and
$\mathbb{C}^{d_j}$ the representation space. In turns, from
Eq.~(\ref{eq:invariant-form}), Schur-Weyl duality gives the form of a
generic permutation invariant operator $X$ on $(\mathbb{C}^2)^{\otimes
  M}$:
\begin{equation}
[X,\Pi_\sigma^M]=0\Longleftrightarrow X=\bigoplus_{j=j_0}^{M/2}X^j\otimes I_{d_j},
\label{eq:schur-weyl-form}\end{equation}
where $X^j$ is an operator on $\mathbb{C}^{2j+1}$.
\subsubsection{Decomposition of many-copies qubit states}
As an application of Schur-Weyl duality, let's consider the
decomposition of the many-copies qubit states $\rho^{\otimes N}$. This
decomposition has been first given in Ref.~\cite{many-copies-decomp}.
For a complete and detailed proof see Ref.~\cite{superbro-pra}.

Indeed, such many-copies states are invariant under permutations of
single qubit systems. The state $\rho^{\otimes N}$ admits then a
decomposition as in Eq.~(\ref{eq:schur-weyl-form}), explicitly
\begin{equation}\label{eq:many-copies-decomp}
  \rho^{\otimes N}=\left(\frac{1-r^2}{4}\right)^{N/2}\bigoplus_{l=l_0}^{N/2}\sum_{n=-l}^l\left(\frac{1+r}{1-r}\right)^n|l,n\>\<l,n|\otimes I_{d_l},
\end{equation}
where, as usual, $\rho=(I+r\vec k\cdot\vec\sigma)/2$, $\|\vec k\|=1$
and $|l,n\>$ are the eigenvectors of the angular momentum along $\vec
k$, namely $J_{\vec k}^{(l)}$. Notice that Eq.
(\ref{eq:many-copies-decomp}) exhibits a singularity for $r=1$ due to
the particular rearrangement of terms. However, the limit for $r\to 1$
exists finite, as it can be seen from the equivalent expression
\begin{equation}
  \rho^{\otimes N}=\bigoplus_{l=l_0}^{N/2}\left(\frac{1-r^2}{4}\right)^{N/2-l}\sum_{n=-l}^l\left(\frac{1+r}{1-r}\right)^{l+n}|l,n\>\<l,n|\otimes I_{d_l}.
\end{equation}

\section{Optimization in a covariant setting}
Let us consider a family $\sF=\{\rho_\theta\}$ of quantum states of
the input system $\sH$. In most cases of physical interest, such a
family is invariant under the action of a unitary representation $U_g$
on $\sH$ of a group $\gG$, in formula:
\begin{equation}
  U_g\rho_\theta U_g^\dag=\rho_{g(\theta)}\in\sF,\qquad\forall\rho\in\sF,\qquad\forall g\in\gG.
\end{equation}
On such a family of states we are concerned about a particular mapping
$\map M$ of $\sF$ onto another family $\sF'=\{\sigma_\theta\}$ of
states of the output system $\sK$ invariant under the action of
another unitary representation $V_g$ of the \emph{same} group $\gG$.
The mapping $\map M$ can be completely general, even physically non
allowable. Let $\map M$ be covariant, namely, $\map
M(\rho_\theta)=\sigma_{\theta}$.

Whatever $\map M$ is, we introduce a physical channel $\map E$ and a
merit function $\mathfrak{F}$, depending on $\theta$ and $\map M$,
such that $\mathfrak{F}[\map E(\rho_\theta),\sigma_{\theta}]\equiv
\mathfrak{F}(\theta)$ achieves its maximum when $\map
E(\rho_\theta)=\sigma_{\theta}$. In other words, $\mathfrak{F}$
quantifies how well the channel $\map E$ approximates the mapping
$\map M$. Assuming transitive action of $\gG$ on $\sF$, that is,
\begin{equation}
\forall\theta,\ \exists g\in\gG:\ \theta=g(\theta_0)\textrm{ for a fixed }\theta_0,
\end{equation}
a further natural requirement is the invariance property of
$\mathfrak{F}$:
\begin{equation}\label{eq:covariant-merit}
  \mathfrak{F}(g(\theta_0))=\mathfrak{F}\left[\map E(U_g\rho_{\theta_0}U_g^\dag),V_{g}\sigma_{\theta_0}V_{g}^\dag\right]=\mathfrak{F}\left[V_{g}^\dag\map E(U_g\rho_{\theta_0}U_g^\dag)V_{g},\sigma_{\theta_0}\right]=\mathfrak{F}(\theta_0).
\end{equation}
Then the function to be maximized is the average score
\begin{equation}\label{eq:average-merit}
\overline{\mathfrak{F}}=\int_\gG\mathfrak{F}\left[\map
E(\rho_{g(\theta_0)}),\sigma_{g(\theta_0)}\right]\ \d g=\mathfrak{F}(\theta_0).
\end{equation}

The basic point is that, if the optimum average score is reached by
some channel $\map E$, it is always possible to achieve the optimum
also by a covariant channel $\widetilde{\map E}$, namely such that
$\widetilde{\map E}(U_g\rho U_g^\dag)=V_g\widetilde{\map
  E}(\rho)V_g^\dag$. Indeed, from Eqs. (\ref{eq:covariant-merit}) and
(\ref{eq:average-merit}) it turns out that\footnote{We also assume
  $\mathfrak{F}$ linear in the l.~h.~s. slot.}
\begin{equation}
\begin{split}
  \overline{\mathfrak{F}}&=\int_\gG \mathfrak{F}\left[V_g^\dag\map E(\rho_{g(\theta_0)})V_g,\sigma_{\theta_0}\right]\d g\\
  &=\mathfrak{F}\left[\int_\gG V_g^\dag\map E(\rho_{g(\theta_0)})V_g\
    \d
    g,\sigma_{\theta_0}\right]\\
&\equiv\mathfrak{F}\left[\widetilde{\map
      E}(\rho_{g(\theta_0)}),\sigma_{\theta_0}\right],
\end{split}
\end{equation}
where we defined
\begin{equation}
\widetilde{\map E}(\rho_{g(\theta_0)})=\int_\gG
V_g^\dag\map E(\rho_{g(\theta_0)})V_g\ \d g.
\end{equation}
It is simple to verify that $[R_{\widetilde{\map{E}}},V_g\otimes
U_g^*]=0$, namely, $\widetilde{\map E}$ is covariant and, by
construction, it achieves the optimal average score
$\overline{\mathfrak F}$.

Hence, in the following we can restrict the optimization procedure to
covariant channels, which form a convex set.  By introducing
appropriate convex merit functions, we can moreover search for the
optimum channel within the border of the convex set, since convex
functions defined on convex sets achieve their extremal values on the
border. In the cases in which we are able to characterize extremal
covariant channels, we can then explicitly single out channels
optimizing the given merit function.

\section{Universally covariant channels}
Universal covariance means, in literature, covariance under the action
of the group $\mathbb{SU}(d)$. Invariant families of states contain
states with fixed spectrum: the most usual choice is to restrict the
analysis to the set of pure states. Given a channel $\map
E:\sS(\sH)\to\sS(\sK)$, universal covariance reads $\map E(U_g\rho
U_g^\dag)=V_g\map E(\rho)V_g^\dag$ where $U_g$ and $V_g$ are unitary
representations of $\mathbb{SU}(d)$ on $\sH$ and $\sK$ respectively.
In the case of pure input states $|\psi\>$, we will consider as merit
function the \emph{fidelity}, namely,
\begin{equation}\label{eq:score-fidelity}
  \mathfrak{F}[\map E(|\psi\>\<\psi|),|\phi\>\<\phi|]\equiv\Tr\left[|\phi\>\<\phi|\;\map E(|\psi\>\<\psi|)\right]=\Tr\left[\left(|\phi\>\<\phi|\otimes|\psi\>\<\psi|^*\right)\;R_\map E\right];
\end{equation}
in the case of mixed input (qubit) states $\rho=(I+r\sigma_z)/2$, we
will consider the purity (the Bloch vector length\footnote{Sometimes
  the purity is defined to be proportional to the \emph{square} of the
  Bloch vector length: $\Tr[\rho^2]=(1+r^2)/2$.}), namely,
\begin{equation}\label{eq:score-purity}
  \mathfrak{F}[\map E(\rho),z]=\Tr\left[\sigma_z\;\map E(\rho)\right]=\Tr\left[\left(\sigma_z\otimes\rho^*\right)\;R_\map E\right].
\end{equation}
It is clear from the form of score functions (\ref{eq:score-fidelity})
and (\ref{eq:score-purity}) that both are convex (linear) in $R_\map
E$ and invariant (see Eq.~(\ref{eq:covariant-merit})).
\subsection{Optimal universal cloning}\label{subsec:universal-cloning}
In this Subsection we shall basically review Ref. \cite{werner} using
Choi-Jamio\l kowski isomorphism. We can't thoroughly apply the
formalism we developed in the previous Sections because a closed form
for Wedderburn's decomposition of $U_g^{\otimes M}$ representation of
$\mathbb{SU}(d)$ is very complicated. We will follow a somewhat
alternative path, finding \emph{a particular} map maximizing the score
function and satisfying covariance and trace-preservation
conditions\footnote{For uniqueness proof see Ref. \cite{werner}.}.

Quantum cloning of an unknown state $\rho^{\otimes N}\to\rho^{\otimes
  M}$, $M>N$, is impossible \cite{nocloning}. Much literature has then
been devoted to searching for optimal physical approximations of
impossible ideal cloning \cite{cloning}. Two basics assumptions are
made in order to make calculations treatable: such optimal machines
should work \emph{equally well} on all input states, and input states
should be \emph{pure} $\rho=\psi\equiv|\psi\>\<\psi|$. The natural
framework to work within is then the universal covariance. The score
function is taken to be the fidelity between the actual output of the
approximation map $\map C(\psi^{\otimes N})$ and the ideal output
$\psi^{\otimes M}$. In terms of the $R_\map C$ operator:
\begin{equation}
  \mathfrak{F}[\map C(\psi^{\otimes N}),\psi^{\otimes M}]=\Tr\left[\left(\psi^{\otimes M}\otimes(\psi^*)^{\otimes N}\right)\ R_\map C\right],
\end{equation}
where $R_\map C$, in order to satisfy universal covariance of $\map
C$, is such that
\begin{equation}
[R_\map C,U_g^{\otimes M}\otimes(U_g^*)^{\otimes N}]=0.
\end{equation}
Let $P_S^{(N)}=\left(P_S^{(N)}\right)^*$ be the projection over the
totally symmetric subspace $\sH_S$ of the input system
$\sH=(\mathbb{C}^d)^{\otimes N}$.  Since $\psi^{\otimes
  N}P_S^{(N)}=\psi^{\otimes N}$, we have that
\begin{equation}
  \mathfrak{F}[\map C(\psi^{\otimes N}),\psi^{\otimes M}]\le\Tr\left[\left(\psi^{\otimes M}\otimes P_S^{(N)}\right)\ R_\map C\right].
\end{equation}
The channel $\map C$ is universally covariant, whence, from Eq.
(\ref{eq:group-average}),
\begin{equation}
  \map C\left(P_S^{(N)}\right)=\int\d gU_g^{\otimes M}\map C\left(P_S^{(N)}\right)(U_g^\dag)^{\otimes M}=\frac{\Tr\left[\map C\left(P_S^{(N)}\right)\right]}{d[M]}P_S^{(M)}+O,
\end{equation}
where $O$ collects all other contributions coming from partially
symmetric/antisymmetric invariant subspaces, and
$d[M]=\binom{d+M-1}{M}$ is the dimension of the totally symmetric
subspace. Actually, terms in $O$ does not contribute to the fidelity
since $\psi^{\otimes M}$ is a symmetric state\footnote{Here it is
  crucial that $\rho=\psi$ is pure.  Otherwise $\rho^{\otimes M}$
  could also have non-null components on partially
  symmetrized/antisymmetrized subspaces.}, hence, w.~l.~o.~g., we
write
\begin{equation}
\map C\left(P_S^{(N)}\right)=\frac{d[N]}{d[M]}P_S^{(M)},
\end{equation}
and obtain the following upper bound for the score function:
\begin{equation}\label{eq:optimal-univ-cloning-fid}
\mathfrak{F}\le\frac{d[N]}{d[M]}.
\end{equation}
One can easily verify that the positive operator
\begin{equation}
  R_\map C=\frac{d[N]}{d[M]}\left(P_S^{(M)}\otimes I^{\otimes N}\right)\left(I^{\otimes M-N}\otimes|I^{\otimes N}\kk\bb I^{\otimes N}|\right)\left(P_S^{(M)}\otimes I^{\otimes N}\right),
\end{equation}
is invariant, properly normalized to trace-preservation\footnote{In
  the sense that $\Tr_\sK[R_\map C]=I_{\sH_S}$.}, and saturates the
bound (\ref{eq:optimal-univ-cloning-fid}). With a little abuse of
notation, we denoted with $|I^{\otimes N}\kk$ the non-normalized
maximally entangled vector in $(\mathbb{C}^d)^{\otimes 2N}$
\begin{equation}
|I^{\otimes N}\kk=\sum_{i_1,\dots,i_N=0}^{d-1}\underbrace{|i_1\>\otimes\dots\otimes|i_N\>}_{(\mathbb{C}^d)^{\otimes N}}\otimes\underbrace{|i_1\>\otimes\dots\otimes|i_N\>}_{(\mathbb{C}^d)^{\otimes N}},
\end{equation}
such that
\begin{equation}
\Tr\left[\left(\psi^{\otimes N}\otimes(\psi^*)^{\otimes N}\right)\ |I^{\otimes N}\kk\bb I^{\otimes N}|\right]=\Tr[\psi^2]^N=1,
\end{equation}
since $\psi$ is pure. From Choi-Jamio\l kowski inverse formula, one
can verify that the action of the optimal universal cloning is as
given in Ref.~\cite{werner}, that is,
\begin{equation}\label{eq:optimal-univ-cloning-map}
\map C(\psi^{\otimes N})=\frac{d[N]}{d[M]}P_S^{(M)}(I^{\otimes(M-N)}\otimes\psi^{\otimes N})P_S^{(M)}.
\end{equation}

\subsection{Optimal universal \texttt{NOT}-gate}\label{subsec:UNOT}
Another unphysical mapping with a naturally emerging covariant
structure is the \texttt{NOT}-gate. In this Subsection we shall
derive, following Ref.~\cite{peristalsi}, the optimal physical
approximation of the ideal quantum-\texttt{NOT}.  Actually, in
Subsection \ref{subsec:unot-and-uclon}, we shall also show that
optimal cloning and optimal \texttt{NOT} are intimately related.

Let us consider a $d$-dimensional system $\sH$ described by the pure
state $\psi\equiv|\psi\>\<\psi|$. When $d=2$, it makes sense to
consider the \texttt{NOT}-gate, which, generalizing the classical
mapping $0\to 1$ and $1\to 0$, sends an unknown pure state to its
\emph{unique} orthogonal complement. Such orthogonal complement, a
part from a fixed unitary transformation, is the transposition of the
input state. This fact explains why \emph{perfect} \texttt{NOT}-gate
is not physical, since transposition is the simplest example of
positive transformation that is not \emph{completely} positive. In
\cite{qubit-unot} the case $d=2$ is addressed and the optimal
universal approximation is worked out.  Here we generalize the result
for all finite dimensions and pure input states.

First of all, it is clear that for $d>2$ the orthogonal complement of
a pure state is not uniquely defined. Hence we shall construct the map
$\map T$ approximating the transposition, which, on the contrary, is
uniquely defined---once fixed a basis in $\sH$. Universal covariance
for a channel whose output transforms as the transposed input, that
is, $\map T(U_g\rho U_g^\dag)=U_g^*\map T(\rho)U_g^T$, reads, as
usual, as an invariance property for $R_\map T$:
\begin{equation}
[R_\map T,U_g^*\otimes U_g^*]=0.
\end{equation}
The unitary representation $(U_g^*)^{\otimes 2}$ of $\mathbb{SU}(d)$
decomposes the space $\sH^{\otimes 2}$ into the irreducible totally
symmetric and totally antisymmetric subspaces, $\sH^{\otimes 2}_S$ and
$\sH^{\otimes 2}_A$ respectively. Hence $R_\map
T=r_SP_S^{(2)}+r_AP_A^{(2)}$, where $P_{S,A}^{(2)}:\sH^{\otimes
  2}\to\sH^{\otimes 2}_{S,A}$ are orthogonal projections.

The covariant score function $\mathfrak{F}$ is taken to be the
fidelity $\Tr[\psi^*\map T(\psi)]$, as always when dealing with pure
states. From the form of $R_\map T$:
\begin{equation}
  \mathfrak{F}=\Tr[(\psi^*)^{\otimes 2}R_\map T]=r_S,
\end{equation}
and $r_S$ has to be maximized consistently with trace-preservation
condition\\\mbox{$\Tr_\sH[r_SP_S^{(2)}]=I$}. Noticing that
$P_S^{(2)}=(I^{\otimes 2}+S)/2$, where $S$ is the swap-operator
between the two spaces, its partial trace is easily computed as
$\Tr_\sH[r_SP_S^{(2)}]=Ir_S(d+1)/2$. The optimal universal
approximation of the transposition map is then uniquely described by
\begin{equation}\label{eq:optimal-not-R}
  R_\map T=\frac{2}{d+1}P_S^{(2)},
\end{equation}
and it achieves optimal fidelity
\begin{equation}\label{eq:optimal-fid-univ-not}
\mathfrak F=2/(d+1).
\end{equation}
Remarkably, such value for
the fidelity equals the fidelity of optimal state estimation over one
copy \cite{d-dim-state-esteem}. This means that, even if the
optimization has been performed in a general setting, the resulting
channel $\map T$, that is optimal and unique, corresponds to nothing
but a trivial \emph{measure-and-prepare} scheme. In other words,
optimal universal transposition can simply be achieved by performing
the optimal state estimation over one copy---the input copy---and then
preparing the transposed of the \emph{estimated} state. This aspect is
usually referred to as \emph{classicality} of the channel. We will see
in Subsection \ref{subsec:phasenot} that, in the case of multi-phase
covariant transposition, this classical limit can be breached.

\subsection{Universal qubit superbroadcasting}\label{subsec:Usuperbro}
Broadcasting of quantum states is a generalization of cloning, in the
sense that given an unknown input state $\rho\in\sS(\sH)$, the
broadcasting machine $\map B$ is allowed to return a generally
entangled output $\Sigma\in\sS(\sH^{\otimes 2})$ such that
$\Tr_1[\Sigma]=\Tr_2[\Sigma]=\rho$. In \cite{no-broad} it's been
proved that it is not possible to broadcast with the same channel two
noncommuting quantum states. This result is generally referred to as
the \emph{no-broadcasting theorem}. Actually, the proof holds only for
single-copy input state; allowing for multiple-copies input, it is
possible to construct a channel broadcasting a whole invariant family
of states. Moreover, considering as merit function the Bloch vector
length (in the case of qubits, see Eq.  (\ref{eq:score-purity})), the
optimal broadcasting channel actually purifies the input state, in the
sense that the single-site reduced output commutes with the input
(hence their Bloch vectors are parallel) being at the same time purer
(with longer Bloch vector) than the input. We will refer to such a
broadcasting-purifying gate as the \emph{superbroadcaster}
\cite{superbroadcasting}. Of course, the superbroadcaster can be made
a ``perfect'' broadcaster by appropriately mixing the output state
with the maximally chaotic state $I/2$ (this procedure simply
corresponds to a depolarizing channel isotropically shrinking the
Bloch vector towards the center of the Bloch sphere).

In what follows we will explicitly derive such an optimal
superbroadcasting machine by thoroughly using group-theoretical
techniques of Section \ref{sec:group-techniqes}.
\subsubsection{Permutation invariance and universal covariance}
We consider a map $\map B$ taking $N$ copies of an unknown qubit state
$\rho$ to a global output state of $M>N$ qubits. A first natural
requirement is that each final user receives the same reduced output
state\footnote{This requirement alone could not give rise to
  permutation invariant output states. However, it is possible to
  prove that one can always find an optimal map satisfying this
  property, see Ref.~\cite{superbro-pra}.}. This fact, along with the
obvious permutation invariance of the input $\rho^{\otimes N}$, leads
to a Choi-Jamio\l kowski operator that must satifsy the following
invariance property (see Eq.~(\ref{eq:perm-inv-0})):
\begin{equation}\label{eq:perm-inv}
  \left[\Pi_\sigma^M\otimes\Pi_\tau^N,R_\map B\right]=0,\qquad\forall\sigma,\tau,
\end{equation}
where $\Pi_\sigma^M$ and $\Pi_\tau^N$ are (real) representations of
the permutation group of the $M$ output and the $N$ input systems,
respectively. From Eq.~(\ref{eq:schur-weyl-form}) the form of $R_\map
B$ follows
\begin{equation}\label{eq:perm-inv-R}
  R_\map B=\bigoplus_{j=j_0}^{M/2}\bigoplus_{l=l_0}^{N/2}R_{jl}\otimes I_{d_j}\otimes I_{d_l},
\end{equation}
where $R_{jl}$ is an operator on
$\mathbb{C}^{2j+1}\otimes\mathbb{C}^{2l+1}$ and $d_j$ and $d_l$ are
the Clebsch-Gordan multiplicities given in
Eq.~(\ref{eq:CB-multiplicities}).

Eq.~(\ref{eq:perm-inv-R}) takes into account only permutation
invariance of input and output states: it can then be further
specialized to different group-covariances. In this Subsection we will
deal with $\mathbb{SU}(2)$ covariance (see Subsection
\ref{subsec:phasebroad} for $\mathbb{U}(1)$ covariance). According to
Subsection \ref{subsec:sud-covariance}, since
$W_g^*=\sigma_yW_g\sigma_y$, such covariance condition rewrites as
$\left[S_\map B,W_g^{\otimes(M+N)}\right]=0$, where $W_g$ is the
defining representation of $\mathbb{SU}(2)$ and \mbox{$S_\map
  B=(I^{\otimes M}\otimes\sigma_y^{\otimes N})R_\map B(I^{\otimes
    M}\otimes\sigma_y^{\otimes N})$}. Hence $S_\map B$ splits as
\begin{equation}\label{eq:onlyonerunning}
  S_\map B=\bigoplus_{j=j_0}^{M/2}\bigoplus_{l=l_0}^{N/2}\bigoplus_{J=|j-l|}^{j+l}s^J_{j,l}P^J_{j,l}\otimes I_{d_j}\otimes I_{d_l},
\end{equation}
where $P^J_{j,l}$ is the orthogonal projection of the space
$\mathbb{C}^{2j+1}\otimes\mathbb{C}^{2l+1}$ onto the
$J$-representation and satisfies the simple properties:
\begin{equation}
  \Tr[P^J_{j,l}]=2J+1,\qquad\Tr_j[P^J_{j,l}]=\frac{2J+1}{2l+1}I_{2l+1},\qquad\Tr_l[P^J_{j,l}]=\frac{2J+1}{2j+1}I_{2j+1}.
\end{equation}
\subsubsection{Classification of extremal points}
Since $S_\map B$ has to be positive, all $s^J_{j,l}$ are positive real
numbers and trace-preservation condition $\Tr_\sK[S_\map B]=I_\sH$
reads
\begin{equation}
  \Tr_\sK[S_\map B]=\bigoplus_{l=l_0}^{N/2}\sum_{j=j_0}^{M/2}\sum_{J=|j-l|}^{j+l}s^J_{j,l}d_j\frac{2J+1}{2l+1}I_{2l+1}\otimes I_{d_l}=I^{\otimes N}.
\end{equation}
The latter is equivalent to
\begin{equation}
\sum_{j=j_0}^{M/2}\sum_{J=|j-l|}^{j+l}s^J_{j,l}d_j\frac{2J+1}{2l+1}=1,\qquad\forall l.
\end{equation}
To single out optimal maps, here we adopt the Bloch vector length
merit function (\ref{eq:score-purity}). This is a linear merit
function, thus optimal maps lie on the border of the convex set of
covariant channels described by $S_\map B$ operators. The problem is
how to characterize extremal $S_\map B$ operators compatible with
complete positivity and trace-preservation constraints. Since
$S_\map B$ is diagonal in indeces $j$ and $J$, extremal $S_\map B$
operators are classified by functions $j=j_l$ and $J=J_l$, leading to
the following expression for extremal $S_\map B$ operators
\begin{equation}
S_\map B=\bigoplus_{l=l_0}^{N/2}\frac{2l+1}{2J_l+1}P^{J_l}_{j_l,l}\otimes\frac{I_{d_{j_l}}}{d_{j_l}}\otimes I_{d_l}.
\end{equation}
\subsubsection{Optimization}
We now feed the input state $\rho^{\otimes N}$ into the channel.
Since we are working in a universally covariant setting, we can write,
w.~l.~o.~g., $\rho=(I+r\sigma_z)/2$, that is, an input state with
Bloch vector along $z$-axis. The global output state $\Sigma$ is
\begin{equation}
\Sigma=\Tr_\sH[I^{\otimes M}\otimes(\sigma_y\rho^*\sigma_y)^{\otimes N}\ S_\map B]=\Tr_\sH[I^{\otimes M}\otimes{\widetilde\rho}^{\otimes N}\ S_\map B]\,,
\label{maponinp}
\end{equation}
where $\widetilde\rho$ denotes the \texttt{NOT} of $\rho$,
corresponding to the inversion $r\to -r$ (or, equivalently, $r_\pm\to
r_\mp$). By means of the decomposition (\ref{eq:many-copies-decomp})
for $\widetilde\rho^{\otimes N}$, we get
\begin{equation}
  \Sigma=\left(\frac{1-r^2}{4}\right)^{N/2}\sum_{l=l_0}^{N/2}\frac{2l+1}{2J_l+1}\frac{d_l}{d_{j_l}}\sum_{n=-l}^l\left(\frac{1-r}{1+r}\right)^{n}\Tr_l\left[I_{2j_l+1}\otimes|ln\>\<ln|\ P^{J_l}_{j_l,l}\right]\otimes I_{d_{j_l}}\,.
\end{equation}
From the form of the map, it is guaranteed that $\Sigma$ is
permutation invariant. Hence it makes sense to speak about \emph{the}
reduced state $\sigma\equiv\Tr_{M-1}[\Sigma]$, regardless of
\emph{which} particular reduced state. In \cite{superbro-pra} it is
shown that $[\sigma_z,\sigma]=0$, namely, the reduced state Bloch
vector is along $z$-axis. The merit function is then
\begin{equation}
  \mathfrak{F}=\Tr\left[\left(\sigma_z\otimes I^{M-1}\right)\;\Sigma\right]\equiv r',
\end{equation}
where $r'$ is the Bloch vector length of $\sigma$. After a lengthy
calculation (see \cite{superbro-pra}), the optimal channel turns out
to be the one with $j_l=M/2$ and $J_l=M/2-l$, regardless of the number
of input copies and of the spectrum of $\rho$.
\begin{figure}
\centering
\includegraphics[width=14cm]{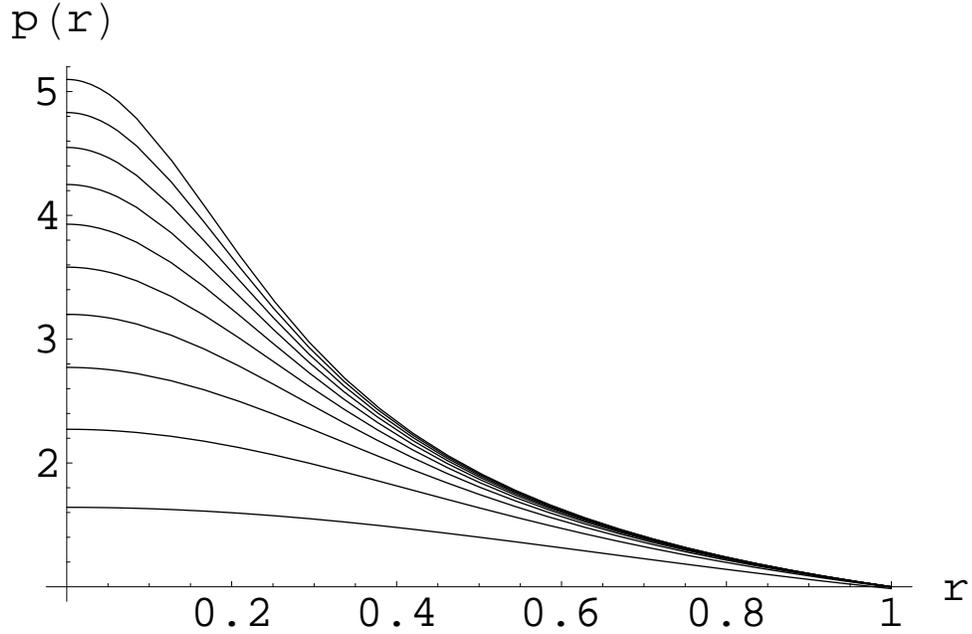}
\caption{The plot shows the behaviour of the scaling factor
  $p^{N,N+1}(r)$ versus $r$, for $N$ ranges from 10 to 100 in steps of
  10, in the universal case. Notice that there is a wide range of
  values of $r$ such that $p^{N,N+1}(r)>1$.}\label{plot:univ-pr-n+1}
\end{figure}
The optimal superbroadcasting achieves the following scaling factor
$p^{N,M}(r)\equiv r'/r$:
\begin{equation}
  p^{N,M}(r)=-\frac{M+2}{Mr}\left(\frac{1-r^2}{4}\right)^{N/2}\sum_{l=l_0}^{N/2}\frac{d_l}{l+1}\sum_{n=-l}^l n\left(\frac{1-r}{1+r}\right)^{n}\,.
\end{equation}
The two limiting cases are $M=N+1$
\begin{equation}
  p^{N,N+1}(r)=-\frac{N+3}{(N+1)r}\left(\frac{1-r^2}{4}\right)^{N/2}\sum_{l=l_0}^{N/2}\frac{d_l}{l+1}\sum_{n=-l}^l n \left(\frac{1-r}{1+r}\right)^{n},
\end{equation}
and $M=\infty$
\begin{equation}
  p^{N,\infty}(r)=-\frac1r\left(\frac{1-r^2}{4}\right)^{N/2}\sum_{l=l_0}^{N/2}\frac{d_l}{l+1}\sum_{n=-l}^l
  n\left(\frac{1-r}{1+r}\right)^{n}.
\end{equation}
By plotting scaling factors for different values of $N$ and $M$, it
turns out that, in the universal case, superbroadcasting first emerges
for $N=4$ (in Subsection \ref{subsec:phasebroad} we will see that, in
the phase-covariant case, superbroadcasting first emerges for $N=3$).
Quite surprisingly, for a sufficiently large number of input copies
($N\ge 6$) it is possible to superbroadcast quantum states even to an
infinte number of receivers. In Fig. \ref{plot:univ-pr-n+1} there are
the plots of $p^{N,N+1}(r)$ for $10\le N\le 100$ in steps of 10.
Notice that for $r\to 1$ all curves go below one: indeed optimal
universal cloning of pure states never achieves fidelity one, see Eq.
(\ref{eq:optimal-univ-cloning-fid}).

\begin{figure}
\centering
\includegraphics[width=14cm]{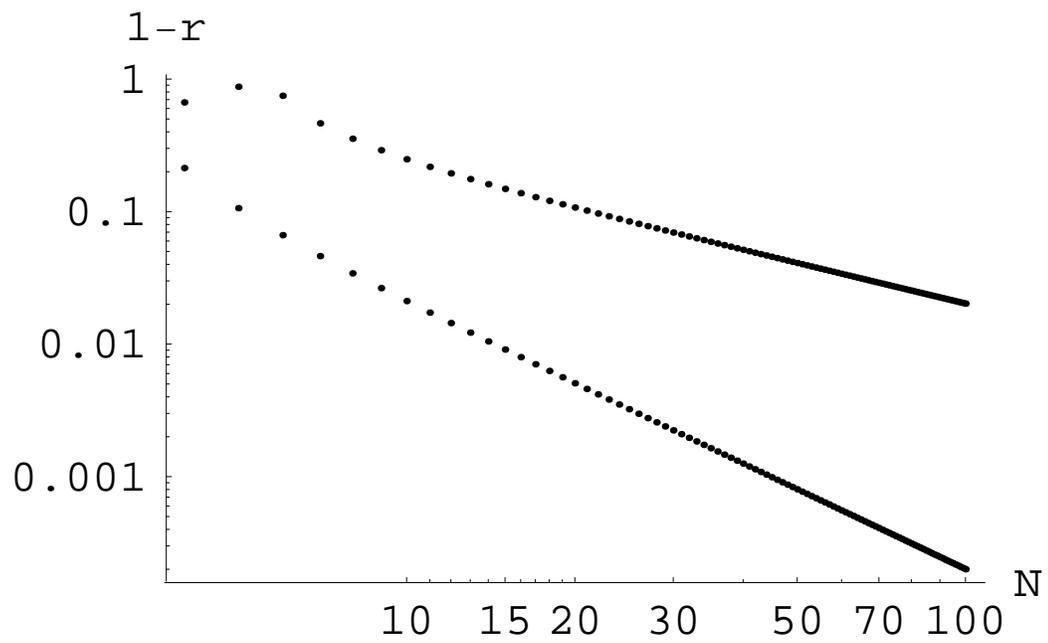}
\caption{Logarithmic plot of $(1-r^*)$ versus $N$ in the universal
  case. The upper line refers to the case $M=\infty$ and shows a
  behaviour like $1/N$. The lower line is for $M=N+1$ and goes like
  $2/N^2$.}\label{plot:univ-rstar}
\end{figure}

A compact way to describe the performances of the optimal
superbroadcaster is to introduce the parameter $r^*$, implicitly
defined by the equation
\begin{equation}
p^{N,M}(r^*)=1.
\end{equation}
Clearly, $r^*$ actually depends on $N$ and $M$. By the monotonicity of
$p(r)$, for $r<r^*$ there is superbroadcasting. Hence, $r^*>0$ means
that superbroadcasting is possible. As we already said, for $N\ge 6$,
$r^*>0$ for all $M$. For $N=5$, $r^*>0$ for $M\le 21$. For $N=4$,
$r^*>0$ for $M\le 7$. Moreover, as $N$ and $M$ get closer, $r^*\to 1$,
as expected. In Fig.~\ref{plot:univ-rstar} there are the plots of
$1-r^*(N,M)$, for $M=N+1$ and $M=\infty$. With good approximation, the
two curves have power law $1-r^*(N,N+1)\propto 2/N^2$ and
$1-r^*(N,\infty)\propto 1/N$.
\section{Phase-covariant channels}
Multi-phase rotations in $d$ dimensions, see Eq.
(\ref{eq:multi-phase-rotation}), obviously form normal subgroups of
$\mathbb{SU}(d)$. In other words, multi-phase covariance group is
``smaller'' than $\mathbb{SU}(d)$ and, consequently, multi-phase
invariant families of states contain ``less states'' than universally
invariant families. Actually, multi-phase invariant families directly
generalize in higher dimension the idea of the equator of the qubits
Bloch sphere\footnote{This idea can be made more rigorous noticing
  that, when $d+1$ mutually unbiased basis can be written, $d$ of them
  are connected by multi-phase rotations, as it happens for qubits.
  See Ref.~\cite{mub}.}. Quite intuitively then, optimization in a
multi-phase covariant setting should generally achieve better
performances than the analogous universal optimization, since the
group is smaller and leaves margin to sharperly tune free parameters.
In what follows we will consider the same examples of the previous
Section (cloning, \texttt{NOT}-gate, and superbroadcasting) in a
multi-phase covariant framework and we will compare the results.

\subsection{Optimal phase-covariant
  cloning}\label{subsec:phasecloning}
The task is to optimally approximate the impossible cloning
transformation $\psi^{\otimes N}\to\psi^{\otimes M}$, where $\psi$ is
an unknown pure state belonging to a family invariant under the
transitive action of the multi-phase group, whose defining
representation is
\begin{equation}
U_\vec\phi=|0\>\<0|+\sum_{n=1}^{d-1}e^{i\phi_n}|n\>\<n|
\end{equation}
(with respect to Eq. (\ref{eq:multi-phase-rotation}) here we put
$\phi_0\equiv 0$, since an overall phase is irrelevant). As before,
since we deal with pure states, the input space $\sH$ is considered to
be the totally symmetric subspace $(\mathbb{C}^d)^{\otimes N}_S$.
Analogously, the output space is $\sK=(\mathbb{C}^d)^{\otimes M}_S$.
Invariant figures of merit are the usual (global) fidelity
\begin{equation}\label{eq:phase-clo-merit}
  \mathfrak{F}_g\left[\map C(\psi_0^{\otimes N}),\psi_0^{\otimes M}\right]=\Tr\left[\map C(\psi_0^{\otimes N})\ \psi_0^{\otimes M}\right],
\end{equation}
and the single-site fidelity 
\begin{equation}
  \mathfrak{F}_s\left[\Tr_{M-1}\left[\map C(\psi_0^{\otimes N})\right],\psi_0\right]=\Tr\left[\map C(\psi_0^{\otimes N})\ \left(\psi_0\otimes I^{\otimes(M-1)}\right)\right],
\end{equation}
where $\psi_0=d^{-1/2}\sum_{i=0}^{d-1}|i\>$ is a fixed state whose
orbit spans all possible input states family. We will adopt $\mathfrak
F_s$, nonetheless, in Ref.~\cite{pheconclon} we proved that
multi-phase covariant cloning maps optimizing single-site fidelity
optimize global fidelity as well. Clearly, it is understood that the
channel $\map C$ satisfies the covariance property
\begin{equation}
  \left[R_\map C,U_\vec\phi^{\otimes M}\otimes (U_\vec\phi^*)^{\otimes N}\right]=0,\qquad\forall\vec\phi,
\end{equation}
so that Eq. (\ref{eq:phase-clo-merit}) makes sense. Last condition
leads to the following form for $R_\map C$
\begin{equation}\label{eq:R-pure-phase-cloning}
R_\map C=\sum_{\{m_j\}}\sum_{\{n'_i\},\{n''_i\}}r^{\{m_j\}}_{\{n'_i\},\{n''_i\}}|\{m_j\}+\{n'_i\}\>\<\{m_j\}+\{n''_i\}|\otimes|\{n'_i\}\>\<\{n''_i\}|,
\end{equation}
where we used the compact notation defined in Eq.
(\ref{eq:symmetric-vectors}). As usual, $R_\map C$ has to be positive,
in order to guarantee complete positivity of $\map C$, and satisfy
trace-preservation condition $\Tr_\sK[R_\map C]=I_\sH$.

\begin{figure}
\centering
\includegraphics[width=14cm]{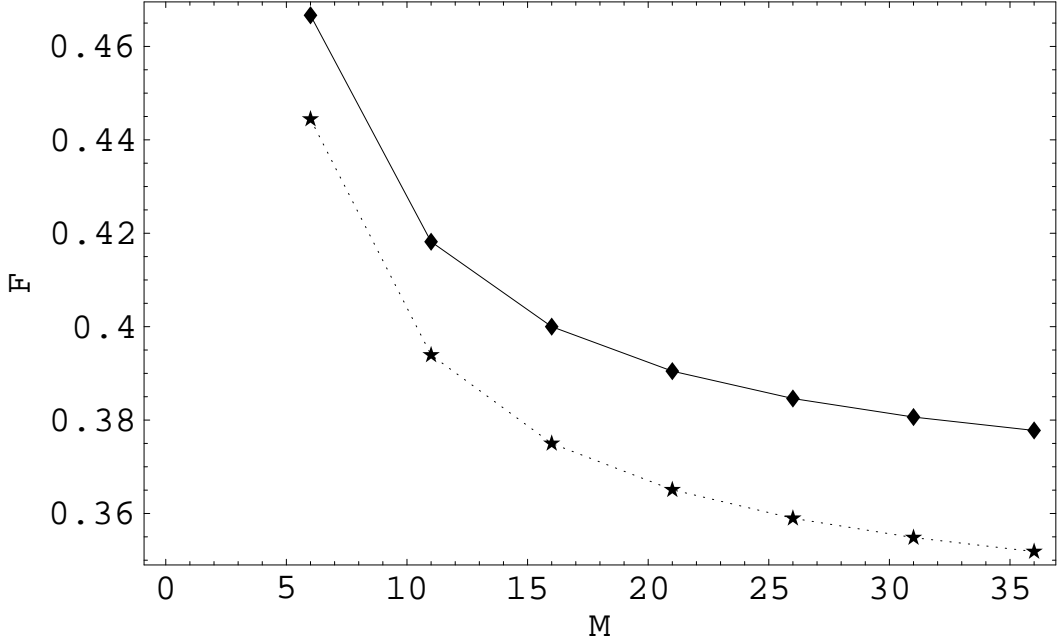}
\caption{Single-site fidelity for $1\to M$ cloning in dimension $d=5$:
  multi-phase (continuous line) and universal (dotted
  line).}\label{fig:phase-clon-pure}
\end{figure}

After lengthy calculations (see Ref.~\cite{pheconclon}), the optimal
multi-phase covariant cloning machine is found to be the one described
by the positive rank-one operator
\begin{equation}\label{eq:R-phase-cov-clon}
  R_\map C=\sum_{\{n_i\},\{n'_i\}}|\{n_i+k\}\>\<\{n'_i+k\}|\otimes|\{n_i\}\>\<\{n'_i\}|,
\end{equation}
where $k$ is a positive integer such that $\sum_i(n_i+k)=M$, hence
equal to $(M-N)/d$. Optimal single-site fidelity is
\begin{equation}\label{eq:opt-fid-pure-phase-clon}
  \mathfrak{F}_s(N,M)=\frac{1}{d}+\frac{1}{Md^{N+1}}\sum_{{\{n_j\} \atop \sum n_j=N-1}}\sum_{i\neq j}\frac{N!}{n_0!\dots n_i!\dots n_j!\dots}\sqrt{\frac{(n_i+k+1)(n_j+k+1)}{(n_i+1)(n_j+1)}},
\end{equation}
which for $N=1$ simplifies to
\begin{equation}
\mathfrak{F}_s(1,M)=\frac{1}{d}+\frac{(d-1)(M+d-1)}{Md^2}.
\end{equation}
In Fig. \ref{fig:phase-clon-pure} there are the plots versus $M$ of
optimal $1\to M$ single-site fidelity in the cases of multi-phase
covariant and universal cloning for $d=5$. Multi-phase covariant
cloning achieves better fidelity than the universal one, as expected.

Notice that our analysis is not completely general because of the
restricting relation that must hold between input and output number of
quantum systems involved
\begin{equation}
  M=N+kd,\qquad k\in\mathbb{N}.
\end{equation}
However it is the most general result on multi-phase covariant cloning
machines described in the literature by now.

\subsection{Optimal phase-covariant
  \texttt{NOT}-gate}\label{subsec:phasenot}
The multi-phase covariant approach to approximate the
\texttt{NOT}-gate is one of the examples in which the performances
improvement, with respect to the universal case, is more apparent.
The transformation we consider is the \texttt{NOT}-gate
$\psi\to\psi^*$ for pure $d$-dimensional states belonging to a
multi-phase invariant family spanned as before by the multi-phase
rotations group $U_\vec\phi$ applied to a fixed seed state
$\psi_0=d^{-1/2}\sum_i|i\>$. The covariant figure of merit is the
fidelity
\begin{equation}
  \mathfrak F\left[\map T(\psi_0),\psi_0^*\right]=\Tr\left[\map T(\psi_0)\ \psi_0\right],
\end{equation}
since $\psi_0^*=\psi_0$ (with the appropriate choice of basis). The
channel $\map T$ must satisfy the covariance property
\begin{equation}
  \left[R_\map
    T,U_\vec\phi^*\otimes U_\vec\phi^*\right]=0.
\end{equation}
The group is abelian so that all irreps are one-dimensional.
Equivalence classes with respective characters are classified in Table
\ref{table1}.
\begin{table}[h]
\begin{center}
\begin{tabular}{c|c}
  Equivalence Classes & Characters\\
  \hline \hline
  $|0\>\otimes|0\>$ & 1\\
  $|1\>\otimes|1\>$ & $e^{-2i\phi_1}$\\
  $\vdots$ & $\vdots$\\
  $|i\>\otimes|i\>$ & $e^{-2i\phi_i}$\\
  $\vdots$ & $\vdots$\\
  $|0\>\otimes|1\>,|1\>\otimes|0\>$ & $e^{-i\phi_1}$\\
  $\vdots$ & $\vdots$\\
  $|i\>\otimes|j\>,|j\>\otimes|i\>,\quad i>j$ & $e^{-i(\phi_i+\phi_j)},\quad i>j$\\
  $\vdots$ & $\vdots$\\
\end{tabular}
\end{center}
\caption{Equivalence classes and respective characters of irreducible
  one-dimensional representations of $(U_\vec\phi^*)^{\otimes 2}$.\label{table1}}
\end{table}
The $R_\map T$ operator then splits into a direct-sum
\begin{equation}
R=\bigoplus_iR_{ii}\bigoplus_{i>j}R_{ij}
\end{equation}
of $1\times 1$ blocks $R_{ii}$ acting on $\Span\{|i\>\otimes|i\>\}$
and $2\times 2$ blocks $R_{ij}$ acting
on\\$\Span\{|i\>\otimes|j\>,|j\>\otimes|i\>\}$.

In Ref.~\cite{phaseconj} there is the complete derivation of the final
form of optimal $R_\map T$ operator as
\begin{equation}
R_\map T=\sum_{i>j}b_{ij}(|ij\>+|ji\>)(\<ij|+\<ji|),
\end{equation}
where $b_{ij}\ge 0$ are matrix elements of a null-diagonal symmetric
bistochastic\footnote{A matrix is called \emph{bistochastic} if all
  its rows and columns entries sum up to one \cite{bhatia}.} matrix.
For $d=2,3$ this constraint suffices to single out a unique optimal
$\map T$, since the only null-diagonal symmetric bistochastic matrix
for $d=2$ is
\begin{equation}
\{b_{ij}\}=\begin{pmatrix}
  0 & 1\\
  1 & 0
\end{pmatrix},
\end{equation}
and for $d=3$
\begin{equation}
\{b_{ij}\}=\begin{pmatrix}
0 & 1/2 & 1/2\\
1/2 & 0 & 1/2\\
1/2 & 1/2 & 0
\end{pmatrix}.
\end{equation}
Already for $d=4$, there are two free parameters $0\le p_1\le 1$ and
$0\le p_2\le 1-p_1$ in defining a null-diagonal symmetric bistochastic
matrix
\begin{equation}\label{eq:NSB-dim4}
\{b_{ij}\}=\begin{pmatrix}
0 & p_1 & p_2 & 1-p_{12}\\
p_1 & 0 & 1-p_{12} & p_2\\
p_2 & 1-p_{12} & 0 & p_1\\
1-p_{12} & p_2 & p_1 & 0
\end{pmatrix},\quad p_{12}=p_1+p_2.
\end{equation}

The achieved optimal fidelity is
\begin{equation}
\mathfrak{F}=\frac{2}{d},
\end{equation}
strictly greater than in the universal case
(\ref{eq:optimal-fid-univ-not}), for all $d$. Moreover, it is
interesting to notice that $2/d$ is also greater than the fidelity of
optimal multi-phase estimation over one copy, derived in
Ref.~\cite{chiara-phaseest} to be $(2d-1)/d^2$.  This means that,
contrarily to the universal case for which the optimal
\texttt{NOT}-gate is classical (see final remarks in
Subsection~\ref{subsec:UNOT}), the multi-phase covariant analogue
breaches the classical limit. The result is particularly striking in
the case of qubits for which it is not possible to perfectly estimate
the phase with finite resources, while it is possible to
\emph{perfectly}---with unit fidelity---transpose an unknown pure
equatorial state by means of a fixed unitary transformation.

\subsection{Phase-covariant qubit
  superbroadcasting}\label{subsec:phasebroad}
In the phase-covariant version of superbroadcasting, we specialize the
permutation invariant form (\ref{eq:perm-inv-R}) imposing the further
constraint
\begin{equation}\label{eq:phase-inv-R}
  \left[R_\map B,U_\phi^{\otimes M}\otimes(U_\phi^*)^{\otimes N}\right]=0.
\end{equation}
Let us now suppose that input states lie on an equator of the Bloch
sphere, say $xy$-plane. Then, $U_\phi$ are precisely rotations along
$z$-axis, namely
\begin{equation}
U_\phi=e^{i\frac{\phi}{2}\sigma_z},
\end{equation}
and invariance condition (\ref{eq:phase-inv-R}) rewrites as
\begin{equation}\label{eq:Rjl-covariance}
  \left[R_{jl},e^{i\phi J^{(j)}_z}\otimes e^{-i\phi J^{(l)}_z}\right]=0,\qquad\forall j,l,
\end{equation}
where $J^{(l)}_z=\sum_{n=-l}^ln|l,n\>\<l,n|$ is the angular momentum
component along $z$-axis in the $l$ representation. A convenient way
to write operators $R_{jl}$ satisfying Eq.~(\ref{eq:Rjl-covariance})
is the following:
\begin{equation}\label{eq:Rjl1}
R_{jl}=\sum_{n=-l}^{l}\sum_{n'=-l}^{l}\sum_{k=l-j}^{j-l}r_{n,n',k}^{jl}|j,n+k\>\<j,n'+k|\otimes|l,n\>\<l,n'|,
\end{equation}
when $j\geq l$, and
\begin{equation}\label{eq:Rjl2}
R_{jl}=\sum_{m=-j}^{j}\sum_{m'=-j}^{j}\sum_{k=j-l}^{l-j}r_{m,m',k}^{jl}|j,m\>\<j,m'|\otimes|l,m+k\>\<l,m'+k|,
\end{equation}
when $j<l$, both expressions exhibiting similar structure as in Eq.
(\ref{eq:R-pure-phase-cloning}). Notice that there are two more
running indeces with respect to the universal case
(\ref{eq:onlyonerunning}). While the index $n'$ in Eq.
(\ref{eq:Rjl1}) simply allows for off-diagonal contributions, the
index $k$ labelling equivalence classes is related to the direction of
the reduced output state Bloch vector, as we will see. In particular
we will show that, in order to get an equatorial output, operators
$R_{jl}$ have to be symmetric in $k$, in the sense that
$r_{n,n',k}^{jl}=r_{n,n',-k}^{jl}$.

\subsubsection{Classification of extremal points and $k$-symmetry}
Trace-preservation now reads
\begin{equation}\label{phase-cpt}
\sum_j\sum_kr_{n,n,k}^{jl}d_j=1,\qquad\forall l,n,
\end{equation}
and, analogously to the universal case, the fact that
$r_{n,n,k}^{jl}\ge 0$ and $R_{jl}$ operators are diagonal with respect
to indices $j$'s and $k$'s implies that extremal points are classified
by functions
\begin{equation}
j=j_l,\qquad k=k_l.
\end{equation}
Equivalently, extremal $R_{jl}$ are proportional to correlation
matrices\footnote{Correlation matrices are positive semi-definite
  matrices with diagonal entries all equal to one.} since they are
positive matrices with diagonal entries $r^{j_l,l}_{n,n,k_l}$ all
equal to $1/d_{j_l}$ (see Eq.  (\ref{phase-cpt})), and extremal
correlation matrices are known in literature \cite{li-tam}.  In
particular, rank-one correlation matrices are extremal, hence rank-one
operators $R_{jl}$ are extremal.

In order to further simplify the general form of $R_\map B$ in Eqs.
(\ref{eq:Rjl1}) and (\ref{eq:Rjl2}), we now impose on the single-site
reduced output state the following additional constraint
\begin{equation}\label{eq:equator}
\Tr_{M-1}\left[\mathcal{B}\left(\frac{I^{\otimes N}}{2^N}\right)\right]=\frac{I}{2}.
\end{equation}
We will see that constraint~(\ref{eq:equator}), on one hand, does not
cause a loss of generality since it does not affect optimality, and,
on the other, clarifies the geometrical interpretation we mentioned
about $k$-indexed degrees of freedom of phase-covariant broadcasting
maps.  In fact we have (for explicit calculation see
Ref.~\cite{superbro-pra})
\begin{equation}
\begin{split}
  \Tr_{M-1}\left[\mathcal{B}\left(\frac{I^{\otimes N}}{2^N}\right)\right]&=\Tr_{M-1}\left[\Tr_\sH\left[\left(I^{\otimes M}\otimes\frac{I^{\otimes N}}{2^N}\right)\ R_\map B\right]\right]\\
  &=\sum_l(2l+1)\frac{d_l}{2^N}\left(\frac{I}{2}+\frac{k_l}{M}\sigma_z\right).
\end{split}
\end{equation}
Since $\sum_l(2l+1)d_l=2^N$, the only condition for
Eq.~(\ref{eq:equator}) is that
\begin{equation}
\sum_l(2l+1)\frac{d_l}{2^N}\frac{k_l}{M}\sigma_z=0.
\end{equation}
In a sense, index $k$ labels a ``tilt'' of the reduced output state
Bloch vector with respect to the equatorial plane. Our requirement is
then a ``null tilt-requirement'', or, in other words, an ``equatorial
output state-requirement'', and it can always be achieved by equally
mixing two extremal maps---generally losing extremality---, the first
labelled by a function $k=\bar k_l$, the second by $k=-\bar k_l$.
\begin{figure}
\centering
\includegraphics[width=14cm]{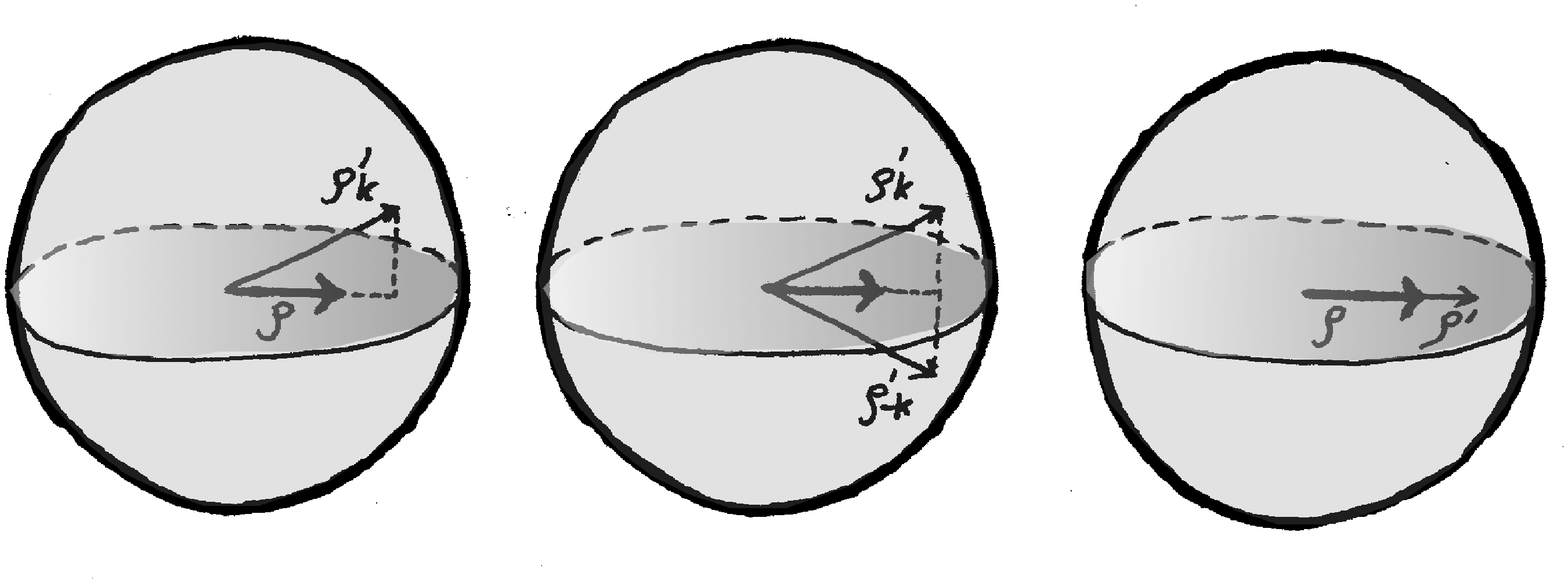}
\caption{Schematic sketch of the $k$-symmetrization
  procedure.}\label{fig:ksymmetry}
\end{figure}
We will refer to such a property as $k$-symmetry property of
bradcasting maps and we showed that $k$-symmetry property is
equivalent to the property of mapping equatorial states to equatorial
states. Notice that a $k$-symmetric map is such that
$r_{n,n',k}^{jl}=r_{n,n',-k}^{jl}$. The strategy to obtain
broadcasting maps optimizing the reduced output state Bloch vector
length is then to search for optimal maps within extremal maps and,
once found the best one, to force $k$-symmetry on it. The procedure is
shown in Fig.~\ref{fig:ksymmetry}. On the left there are the
equatorial mixed input state $\rho$ and the single-site reduced output
$\rho'_k$.  Suppose such an output comes from an extremal map $\map
B_k$ described by $r_{n,n',k}^{jl}$ elements. Notice that, by
covariance, the projection of $\rho'_k$ onto the equator is parallel
with $\rho$.  Consider now another map $\map B_{-k}$, whose elements
are equal to $\tilde r_{n,n',k}^{jl}=r_{n,n',-k}^{jl}$.  Clearly,
$\map B_{-k}$ is a proper channel obeying all covariance and
extremality constraints as $\map B_k$. The output of $\map B_{-k}$ is
in sketched in the middle figure as $\rho'_{-k}$. In order to have an
equatorial output, we mix $\map B_k$ and $\map B_{-k}$ obtaining $\map
B=(\map B_k+\map B_{-k})/2$ whose output
$\rho'=(\rho'_k+\rho'_{-k})/2$, by linearity, lies on the equator (see
the picture on the right). Of course $\map B$ is no more extremal, by
construction. However, the figure of merit we are considering, namely,
the length of the projection of the output Bloch vector onto the
original one, does not change. In this sense, imposing $k$-symmetry
does not affect optimality. Moreover, it is possible to prove that the
$k$-symmetrized output $\rho'$ has higher fidelity with the input
$\rho$ (see Ref.~\cite{superbro-pra})) than the tilted $\rho'_k$ and
$\rho'_{-k}$.

\subsubsection{Optimization}
In Ref~\cite{superbro-pra} it is proved that the channel optimizing
the merit function
\begin{equation}
  \mathfrak{F}=\Tr\left[\left(\sigma_x\otimes I^{M-1}\right)\ \Sigma\right],
\end{equation}
for $x$-oriented input states $\rho=(I+r\sigma_x)/2$, has $j_l=M/2$
for all $l$, and $k_l=0$, for $M-N$ even, while $k_l=\pm 1/2$ for
$M-N$ odd. Hence, for $M-N$ even the optimal superbroadcaster is
already $k$-symmetrized, whereas for $M-N$ odd we must equally mix the
channels coming from $k_l=1/2$ and $k_l=-1/2$. In both cases,
$r_{n,n',k_l}^{j_l,l}=1/d_{j_l}$, for all $n,n',l$. At the end, the
structure of the map $\map B$ depends only on the parity of $M-N$, and
not on the spectrum of $\rho$.

\begin{figure}
\centering
\includegraphics[width=14cm]{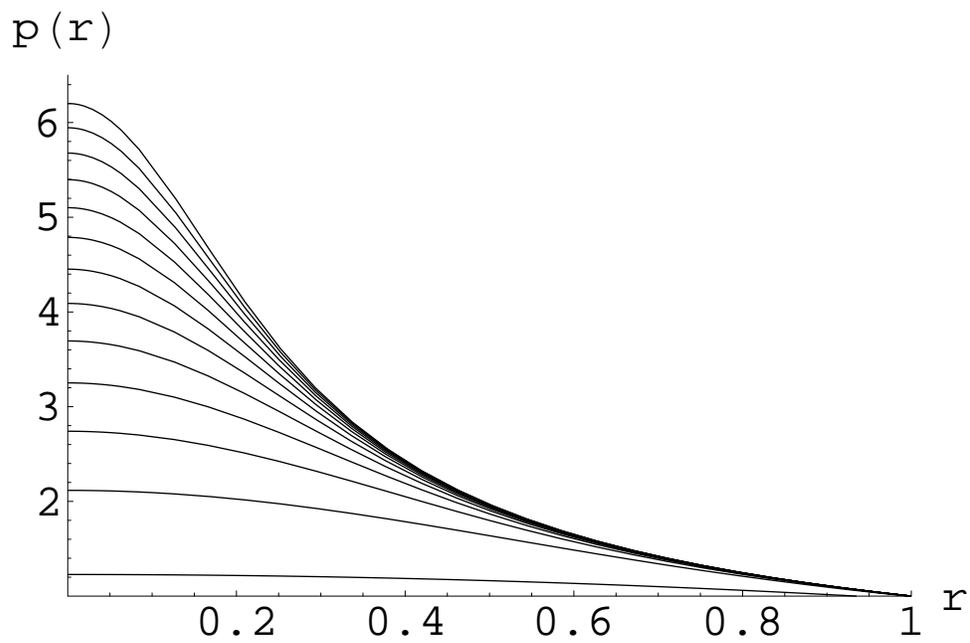}
\caption{The plot shows the behaviour of the scaling factor
  $p^{N,N+1}(r)$ versus $r$, for $N$ ranges from 4 to 100 in steps of
  8, in the phase-covariant case. Notice that there is a wide range of
  values of $r$ such that $p^{N,N+1}(r)>1$.}\label{plot:phase-pr-n+1}
\end{figure}

For $M-N$ even, the optimum scaling
factor $p^{N,M}(r)$ is given by
\begin{equation}\label{eq:phase-pr1}
  p^{N,M}_e(r)=\frac{4}{Mr}\left(\frac{1-r^2}{4}\right)^{N/2}\sum_{l=l_0}^{N/2}d_l\sum_{n=-l}^l\left[\exp\left(J_x^{(l)}\log\frac{1+r}{1-r}\right)\right]_{n,n+1}\left[J_x^{(j)}\right]_{n,n+1},
\end{equation}
while, for $M-N$ odd, it is
\begin{equation}\label{eq:phase-pr2}
  p^{N,M}_o(r)=\frac{4}{Mr}\left(\frac{1-r^2}{4}\right)^{N/2}\sum_{l=l_0}^{N/2}d_l\sum_{n=-l}^l\left[\exp\left(J_x^{(l)}\log\frac{1+r}{1-r}\right)\right]_{n,n+1}\left[J_x^{(j)}\right]_{n+1/2,n+3/2}.
\end{equation}

In Fig.~\ref{plot:phase-pr-n+1} there are the plots of $p^{N,N+1}(r)$
for $4\le N\le 100$ in steps of 8. As in the universal case, all
curves, for $r\to 1$, go below one: indeed optimal phase-covariant
cloning of pure states never achieves fidelity one, see Eq.
(\ref{eq:opt-fid-pure-phase-clon}). However, it is possible to see
that phase-covariant superbroadcasting is more efficient than the
universal one: superbroadcasting first emerges for $N=3$ ($N=4$ in the
universal case) and achieves larger values of $p^{N,M}(r)$ for all
$N$, $M$, and $r$.

\begin{figure}
\centering
\includegraphics[width=14cm]{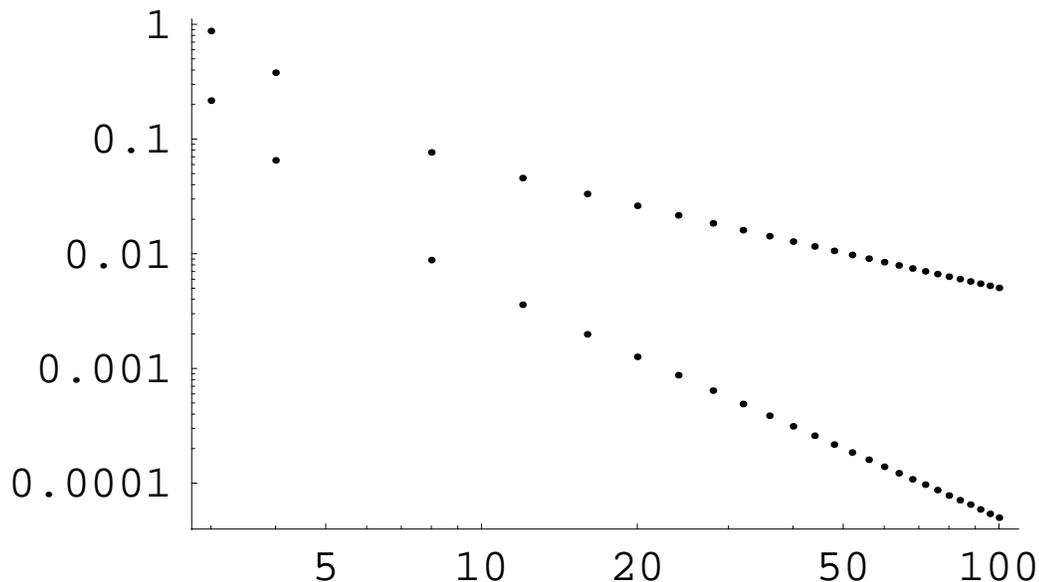}
\caption{Logarithmic plot of $(1-r^*)$ versus $N$ in the
  phase-covariant case. The upper line refers to the case $M=\infty$
  and shows a behaviour like $1/2N$. The lower line is for $M=N+1$ and
  goes like $2/(3N^2)$.}\label{plot:phase-rstar}
\end{figure}

In Fig.~\ref{plot:phase-rstar} there are the plots of $1-r^*(N,M)$,
for $M=N+1$ and $M=\infty$, as done for the universal
superbroadcaster.  With good approximation, the two curves have power
law $2/3N^2$ and $1/2N$, respectively, namely they go to zero faster
than in the universal case, as expected.

%% file: Chap3.tex
\chapter{Realization of Quantum Devices}\label{ch:realization}
In the previous Chapter we explicitly wrote quantum operations coming
out from an optimization procedure in a covariant setting. We gave
such channels in terms of their Choi-Jamio\l kowski operators
(\ref{eq:choi-jam}). However, Choi-Jamio\l kowski representation for
quantum channels, even if very useful in dealing with semi-definite
programming problems, turns out to be quite far from giving the
physical ``recipe'' needed to realize the channel in a laboratory. In
the following we will describe how to unitarily implement a given
quantum channel, in terms of a unitary interaction between the system
and an ancilla. In the first Section, we will provide, following
Refs.~\cite{laurea,unitary-real-pra}, a general method to work out a
physical setting realizing a given channel. In the second part of the
Chapter, we will show how this procedure works in the case of some of
the channels discussed in Chapter~\ref{ch:chap2}.
\section{Unitary dilations of a channel}
Let us given a channel $\map E:\sS(\sH)\to\sS(\sH)$\footnote{Here, for
  simplicity we disregard the case of channels from states on a system
  $\sH$ to states on another system $\sK$, e.~g. the cloning channel
  from $\sS(\sH^{\otimes N})$ to $\sS(\sH^{\otimes M})$. This case can
  be taken into account, see Ref.~\cite{unitary-real-pra} for a more
  general approach.}. The task of this Section is to find an ancilla
system $\sA$, an ancilla pure state $|0\>$, and a unitary operator $U$
on $\sH\otimes\sA$, such that
\begin{equation}
  \map E(\rho)=\Tr_\sA\left[U(\rho\otimes|0\>\<0|)U^\dag\right],
\end{equation}
for all $\rho\in\sS(\sH)$.
\subsection{Stinespring dilation}
Given a channel $\map E$, the Stinespring representation
\cite{stinespring} $(V,\sA)$ of $\map E$ is a kind of ``purification''
of the channel $\map E$, i.~e.
\begin{equation}\label{eq:stinespring}
  \map E(\rho)=\Tr_\sA\left[V\rho V^\dag\right],
\end{equation}
where $V$ is an isometry, i.~e. $V^\dag V=I$, from $\sH$ to
$\sH\otimes\sA$. The Stinespring representation is usually given for
the dual channel (see Subsection~\ref{subsec:quantum-operations})
\mbox{$\dual{\map E}:\sB(\sH)\to\sB(\sH)$} as
\begin{equation}
  \dual{\map E}(O)=V^\dag(O\otimes I_\sA)V.
\end{equation}

Let $\map E(\rho)=\sum_iE_i\rho E_i^\dag$ be a Kraus representation
for $\map E$. Consider now the operators from $\sH$ to $\sH\otimes\sA$
defined as $E_i\otimes|i\>$, where $|i\>$ belongs to a set of
orthonormal vectors in $\sA$. The only trivial condition $\sA$ must
satisfy is $\dim\sA\ge\sharp\{E_i\}$. Then, the sum
\begin{equation}\label{eq:stine-isom}
V=\sum_iE_i\otimes|i\>
\end{equation}
is an isometry, since $V^\dag V=\sum_iE_i^\dag E_i=I_\sH$, and
realizes the channel $\map E$ as in Eq.~(\ref{eq:stinespring}).
\begin{remark}
  Notice that we did not make any assumption on the particular choice
  for the Kraus representation $\{E_i\}$ used to construct the
  Stinespring isometry $V$ in Eq.~(\ref{eq:stine-isom}). When
  $\{E_i\}$ is the canonical Kraus decomposition and
  $\dim\sA=\sharp\{E_i\}$, we will refer to such $V$ as the
  \emph{canonical Stinespring representation} for $\map E$, which
  clearly is the one minimizing the ancillary resources, i.~e. the
  dimension of the ancilla system, needed to physically implement the
  channel.
\end{remark}
\subsection{Unitary dilation}
From Stinespring form (\ref{eq:stinespring}) the existence of a
unitary interaction $U$ between $\sH$ and $\sA$ realizing the channel
$\map E$ is apparent, since every isometry $V$ from $\sH$ to
$\sH\otimes\sA$ can obviously be written as
\begin{equation}
V=U(I_\sH\otimes|0\>),
\end{equation}
where $U$ is a suitable unitary operator on $\sH\otimes\sA$ and $|0\>$
is a fixed normalized state of $\sA$. Now, $|0\>$ is precisely the
ancilla state such that
\begin{equation}\label{eq:generic-unit-real}
  \map E(\rho)=\Tr_\sA\left[U(\rho\otimes|0\>\<0|)U^\dag\right].
\end{equation}

While the existence of a realization $U$ for every channel is a
well-established fact in the literature \cite{ozawa,kraus-unitary},
the problem of giving \emph{explicitly} such interaction for a given
channel can be very difficult. The general procedure given in
Ref.~\cite{unitary-real-pra} basically relies on a repeated
Gram-Schmidt orthonormalizing algorithm applied to the column vectors
of the Stinespring isometry $V$. In this way we are able to find
additional $\dim\sH\times(\dim\sA-1)$ orthonormal vectors to append to
$V$, completing it to a square matrix whose column vectors form an
orthonormal basis for the composite system $\sH\otimes\sA$, i.~e. to a
unitary operator on $\sH\otimes\sA$. In the following Section, we will
see that, in some fortunate cases, the channels optimized in Chapter
\ref{ch:chap2} admit very simple Stinespring isometries, allowing us
to explicitly write unitary operators realizing such channels in
dimension $d$.
\section{Explicit realizations}
\subsection{Universal \texttt{NOT} and cloning
  gates}\label{subsec:unot-and-uclon}
Let us start from the optimal universal \texttt{NOT}-gate $\map T$
derived in Subsection~\ref{subsec:UNOT}. The channel is completely
described by the positive operator $R_\map T$ in
Eq.~(\ref{eq:optimal-not-R}). In order to write $\map T$ in its
Stinespring form, we first have to obtain a Kraus decomposition for
$\map T$. This can be done by expanding $R_\map T$ (see Section
\ref{sec:choi-jam}) as
\begin{equation}
\begin{split}
  R_\map T&=\frac{2}{d+1}P_S^{(2)}=\frac1{d+1}(I+S)\\
  &=\frac1{d+1}\sum_{m,n=0}^{d-1}(|m\>\<m|\otimes|n\>\<n|+|m\>\<n|\otimes|n\>\<m|)\\
  &=\frac1{2(d+1)}\sum_{m,n=0}^{d-1}(|mn\kk+|nm\kk)(\bb mn|+\bb nm|)\\
  &=\sum_{m,n=0}^{d-1}|M^S_{mn}\kk\bb M^S_{mn}|,
\end{split}
\end{equation}
where
\begin{equation}
  M^S_{mn}=\frac1{\sqrt{2(d+1)}}(|m\>\<n|+|n\>\<m|).
\end{equation}
One possible Kraus decomposition is then given by
\begin{equation}
  \map T(\psi )=\sum_{m,n=0}^{d-1}M^S_{mn}\psi M^S_{mn}.
\end{equation}
A Stinespring isometry $V$ such that $\map
T(\psi)=\Tr_{\sA}\left[V\psi V^\dag\right]$ is then\footnote{Notice
  that this Stinespring isometry is not the one minimizing ancillary
  resources. In fact, it comes from a Kraus decomposition which is not
  the canonical one, since the orthogonality condition,
  $\Tr[M^S_{ij}M^S_{kl}]=0$ for all $\{ij\}\neq\{kl\}$, does not hold.
  However, as we will see in the following, this realization allows a
  very intriguing physical interpretation, see
  Ref.~\cite{extremal-clonings}.}
\begin{equation}
  V=\sum_{m,n=0}^{d-1}M^S_{mn}\otimes|mn\kk_{23},
\label{eq:iso}
\end{equation}
where we chose $\sA\equiv\sH^{\otimes 2}$ as ancilla system.
Summarizing, we wrote the optimal \texttt{NOT}-gate $\map T$ by means
of an isometry $V$ embedding the input system $\sH$ into a composite
tripartite system $\sH\otimes\sH\otimes\sH$, in which the last two
spaces represent the ancilla.

Tracing $V\psi V^\dag$ over the last two spaces, we get the channel
$\map T$. What happens if we trace over different combinations of
spaces? In fact, all three spaces are the same and there is no reason
to consider one of them as the preferred system and the remaining ones
as ancillae. Actually, tracing $V\psi V^\dag$ over the \emph{first}
space, one obtains
\begin{equation}
  \Tr_1\left[V\psi V^\dag\right]=\frac {2}{d+1}P_S^{(2)}(I\otimes\psi)P_S^{(2)},
\end{equation}
namely, the optimal $1\to 2$ universal cloning for pure states (see
Eq.~\ref{eq:optimal-univ-cloning-map}). This means that universal
$1\to 2$ cloning and universal \texttt{NOT}-gate are intimately
related and contextually appear on different branches (spaces) of the
same physical setting. Such a coincidence has been experimentally
exploited for qubits in Ref.~\cite{demartini} and theoretically
analyzed and interpreted in generic dimension in
Ref.~\cite{extremal-clonings}. Moreover, it is possible to prove that
$\Tr_3\left[V\psi V^\dag\right]$ optimally approximate the
transformation $\psi\to\psi^*\otimes\psi$ for pure states. Notice that
the cloning map is basis independent, whilst the transposition map
depends on the choice of the basis, which is reflected by the choice
of the particular Stinespring isometry $V$.

In Ref.~\cite{peristalsi} it is possible to find the explicit
calculation deriving a unitary interaction and an ancilla state
realizing at the same time optimal approximations of universal cloning
and transposition. The unitary operator $U$ on
$(\mathbb{C}^d)^{\otimes 3}$ is
\begin{equation}\label{eq:unitary-cloning}
  U=\sum_{p=0}^{d-1}V_{p,p}\otimes\<p|\<p|+\sum_{{p,q=0\atop p<q}}^{d-1}V_{p,q}^{(S)}\otimes \frac{\<p|\<q|+\<q|\<p|}{\sqrt{2}}+\sum_{{p,q=0\atop p<q}}^{d-1}V_{p,q}^{(A)}\otimes \frac{\<p|\<q|-\<q|\<p|}{\sqrt{2}}
\end{equation}
where the three sets of isometries $\left\{V_{p,p}\right\}$,
$\left\{V_{p,q}^{(S)}\right\}$, and $\left\{V_{p,q}^{(A)}\right\}$
from $\sH$ to $\sH^{\otimes 3}$ are defined as
\begin{equation}
\begin{split}
  &V_{p,p}=\sum_{k=0}^{d-1}|k\>|k\oplus p\>|k\oplus p\>\<k\oplus p|,\\
  &V_{p,q}^{(S)}=\frac{1}{\sqrt{2}}\sum_{k=0}^{d-1}|k\>(|k\oplus
  p\>|k\oplus q\>+|k\oplus q\>|k\oplus p\>)\<k\oplus q|,\\
  &V_{p,q}^{(A)}=\frac{1}{\sqrt{2}}\sum_{k=0}^{d-1}|k\>(|k\oplus
  p\>|k\oplus q\>-|k\oplus q\>|k\oplus p\>)\<k\oplus q|.
\end{split}
\end{equation}
Preparing the ancilla state as
\begin{equation}\label{eq:ancilla-cloning}
  |\phi\kk=\sqrt{\frac 2{d+1}}P_S^{(2)}\sum_{r=0}^{d-1}|0\>|r\>,
\end{equation}
the following identity holds
\begin{equation}
U(\psi\otimes|\phi\kk\bb\phi|)U^\dag=V\psi V^\dag,
\end{equation}
namely, the operator $U$ in Eq.~(\ref{eq:unitary-cloning}) together
with the ancilla state $|\phi\kk$ in Eq.~(\ref{eq:ancilla-cloning})
provide a unitary dilation of the Stinespring isometry $V$ in
Eq.~(\ref{eq:iso}), realizing optimal universal $1\to 2$ cloning as
well as optimal universal transposition, depending on what we trace
out after the interaction.

In the case $d=2$, we obtain the network model for universal qubit
cloning of Ref.~\cite{buzek-network}, with
\begin{equation}
U=
\begin{pmatrix}
1 & 0 & 0 & 0 & 0 & 0 & 0 & 0\\
0 & 0 & 0 & 0 & 0 & 1 & 0 & 0\\
0 & 0 & 0 & 0 & 0 & 0 & 1 & 0\\
0 & 0 & 0 & 0 & 0 & 0 & 0 & 1\\
0 & 0 & 0 & 1 & 0 & 0 & 0 & 0\\
0 & 0 & 1 & 0 & 0 & 0 & 0 & 0\\
0 & 1 & 0 & 0 & 0 & 0 & 0 & 0\\
0 & 0 & 0 & 0 & 1 & 0 & 0 & 0\\
\end{pmatrix},
\end{equation}
and $|\phi\kk=\frac{1}{\sqrt{6}}(2|0\>|0\>+|0\>|1\>+|1\>|0\>)$.
\subsection{Phase-covariant cloning and economical maps}
In Subsection \ref{subsec:phasecloning} we obtained the channel
optimally achieving the multi-phase covariant $N\to M$ cloning
transformation. The optimal channel has been described, as usual, by
giving the corresponding $R_\map C$ operator in
Eq.~(\ref{eq:R-phase-cov-clon}). In the analyzed cases, i.~e. when
$M=N+kd$, where $k\in\mathbb{N}$ and $d$ is the dimension of the
single copy system, $R_\map C$ enjoys the relevent property of being
rank-one. This implies that its canonical Kraus representation
contains only one operator, and, to satisfy trace-preservation
constraint (\ref{eq:trace-pres-with-kraus}), such an operator has to
be an isometry. Therefore, the optimal multi-phase covariant $N\to M$
cloning machine $\map C_{N,M}$, for $M=N+kd$, admits the very simple
expression
\begin{equation}
  \map C_{N,M}(\psi^{\otimes N})=V\psi^{\otimes N}V^\dag,
\end{equation}
where $V:\sH^{\otimes N}\to\sH^{\otimes M}$ is an isometry acting as
follows
\begin{equation}
V|\{n_i\}\>=|\{n_i+k\}\>,
\end{equation}
using the compact notation introduced in
Eq.~(\ref{eq:symmetric-vectors}).

This kind of isometric optimal channels attracted attention in the
recent literature as \emph{economical} transformations
\cite{niugrif,economical-durt,economical-cerf}, in the sense that, in
order to physically implement them, there is no need of discarding
additional resources, i.~e. ancillae. In fact, from the point of view
of Stinespring representation, multi-phase covariant cloning is
realizable as
\begin{equation}
  \map C_{N,M}\left(\psi^{\otimes N}\right)=U\left(\psi^{\otimes N}\otimes|0\>\<0|^{\otimes(M-N)}\right)U^\dag,
\end{equation}
namely, with respect to Eq.~(\ref{eq:generic-unit-real}), there is no
partial trace, and the resources needed are just the $(M-N)$ blank
copies where convariantly distribute the information contained in
$\psi^{\otimes N}$.
\subsection{Phase-covariant \texttt{NOT}-gate}
The optimal multi-phase conjugation map has been derived in Subsection
\ref{subsec:phasenot} to be
\begin{equation}
R_\map T=\sum_{i>j}b_{ij}(|ij\>+|ji\>)(\<ij|+\<ji|),
\end{equation}
where $b_{ij}\ge 0$ are matrix elements of a null-diagonal symmetric
bistochastic (NSB) matrix. In this case the map, for $d>2$, is not
unitary or isometric as in the case of phase-covariant cloning.
Moreover, the fact that in dimension $d\ge 4$ there exists a whole set
of equally optimal maps---in one-to-one correspondence with NSB
matrices---makes the problem of finding a physical realization much
more difficult than in the two examples treated before, where the
optimal map was unique. There are basically two paths one can follow:
the first is to search for the most efficient realization, i.~e. the
one minimizing ancillary resources (in this case we will tipically
single out one particular optimal phase-conjugation map achievable
using less resources than the others); the second is to search for the
most flexible realization, i.~e. the one that spans as many as
possible optimal maps by appropriately varying the ``program'' ancilla
state and/or is more robust against noise (this second kind of
realization will clearly require a higher dimensional ancilla system
to encode a ``fault-tolerant'' program).

A good point to start with is the study of the structure of the set of
optimal phase-conjugation channel, or, equivalently, of the set of NSB
matrices. Such matrices form a convex set\footnote{This is because
  their raws and columns are probability distributions}. On the other
hand, every bistochastic matrix is a convex combination of permutation
matrices---this is the content of the Birkhoff theorem \cite{bhatia}.
The null-diagonal and symmetry constraints, however, force the convex
set of NSB matrices to be strictly contained into the convex
polyhedron of bistochastic matrices. This fact causes the extremal NSB
matrices to eventually lie strictly inside the set of bistochastic
matrices, generally preventing them from being permutations.

The geometrical study of the set of NSB matrices and its extremal
points can shed some light on the unusual feature that there exist
different ``equally optimal'' maps. The problem arises for dimension
at least $d=4$. In this case the decomposition of the matrix
$\{b_{ij}\}$ in Eq.~(\ref{eq:NSB-dim4}) into extremal components is
\begin{equation}\label{ext-dec-4}
\begin{split}
\{b_{ij}\}&=p_1\begin{pmatrix}
0& 1& 0& 0\\
1& 0& 0& 0\\
0& 0& 0& 1\\
0& 0& 1& 0
\end{pmatrix}+p_2\begin{pmatrix}
0& 0& 1& 0\\
0& 0& 0& 1\\
1& 0& 0& 0\\
0& 1& 0& 0
\end{pmatrix}+p_3\begin{pmatrix}
0& 0& 0& 1\\
0& 0& 1& 0\\
0& 1& 0& 0\\
1& 0& 0& 0
\end{pmatrix}\\
&=p_1B^{(1)}+p_2B^{(2)}+p_3B^{(3)},
\end{split}
\end{equation}
where $p_1,p_2,p_3\geq 0$ and $p_1+p_2+p_3=1$. A natural question is
now which optimal maps can be achieved with minimal resources.

More explicitly, for $d=4$, we define three unitaries $U_1$, $U_2$ and
$U_3$ on $\mathbb{C}^4\otimes\mathbb{C}^2$ as
\begin{equation}\label{unitaries-for-4}
U_1=\begin{pmatrix}
T_{10} & T_{32}\vspace{0.2cm}\\
T_{32} & T_{10}
\end{pmatrix},\quad
U_2=\begin{pmatrix}
T_{20} & T_{31}\vspace{0.2cm}\\
T_{31} & T_{20}
\end{pmatrix},\quad
U_3=\begin{pmatrix}
T_{30} & T_{21}\vspace{0.2cm}\\
T_{21} & T_{30}
\end{pmatrix},
\end{equation}
where $T_{ij}=|i\>\<j|+|j\>\<i|$. Each of them realizes an extremal
optimal multi-phase conjugation map (corresponding to $p_k=1$ in Eq.
(\ref{ext-dec-4}) for a given $k$), namely
\begin{equation}\label{ext-unit-4}
  \mathcal{T}^{(k)}_4(\rho)=\sum_{i>j}B^{(k)}_{ij}T_{ij}\rho T_{ij}=\Tr_a[U_k\;(\rho\otimes|0\>\<0|_a)\;U_k^\dag],
\end{equation}
where $|0\>\<0|_a$ is a fixed qubit ancilla state. Hence
\emph{extremal phase-conjugation maps for $d=4$ can be achieved with
  just a control qubit}. Notice that the ancilla must not necessarily
be in a pure state, and the optimal map is equivalently achieved for
diagonal mixed ancilla state $\alpha|0\>\<0|_a+\beta|1\>\<1|_a$. By
adding a control qutrit, we can now choose among any of the optimal
maps using the controlled-unitary operator on
$\mathbb{C}^4\otimes\mathbb{C}^2 \otimes\mathbb{C}^3$
\begin{equation}
U=U_1\otimes|0\>\<0| + U_2\otimes|1\>\<1|
+U_3\otimes|2\>\<2|.
\end{equation}
Any optimal multi-phase conjugation map can now be written as
\begin{equation}\label{unitary-4}
\mathcal{T}_4(\rho)=\Tr_{a,b}\left[U\;\left(\rho\otimes|0\>\<0|_a\otimes\sigma_b\right)\;U^\dag\right]
\end{equation}
where $\sigma_b$ is a generic density matrix on $\mathbb{C}^3$. By
superimposing or mixing the three orthogonal states
$\{|0\>,|1\>,|2\>\}$ of the qutrit we control the weights
$p_1,p_2,p_3$ in Eq. (\ref{ext-dec-4}) via the diagonal entries of
$\sigma_b$. In other words, \emph{using a 6-dimensional ancilla it is
  possible to span the whole set of optimal maps}.

Eqs. (\ref{unitaries-for-4})-(\ref{unitary-4}) can be generalized for
higher even dimensions\footnote{The case of odd dimensions is much
  more complicated and will not be analysed here. The problem with odd
  dimensions is that extremal points of the convex set of NSB matrices
  are not permutations. Hence Birkhoff theorem cannot be applied.},
with
\begin{equation}
\begin{split}
&U_k=\sum_{i,j=0}^{\frac{d}{2}-1}T_{k\oplus 2i\oplus 2j,2i\oplus 2j}\otimes|i\>\<j|,\qquad k=1,\dots,d-1,\\
&U=\sum_{k=1}^{d-1}U_k\otimes|k\>\<k|,\\
&\mathcal{T}^{(k)}_d(\rho)=\Tr_a[U_k\;(\rho\otimes|0\>\<0|_a)\;U_k^\dag],\\
&\mathcal{T}_d(\rho)=\Tr_{a,b}\left[U\;\left(\rho\otimes|0\>\<0|_a\otimes\sigma_b\right)\;U^\dag\right]\\
\end{split}
\end{equation}
where $U_k$'s are unitary operators acting on
$\mathbb{C}^d\otimes\mathbb{C}^{d/2}$, $U$ is a control-unitary
operator on
$\mathbb{C}^d\otimes\mathbb{C}^{d/2}\otimes\mathbb{C}^{d-1}$,
$|0\>\<0|_a$ is a fixed $(d/2)$-dimensional pure state, and $\sigma_b$
is a generic $(d-1)$-dimensional density matrix. \emph{The minimum
  dimension of the ancilla space required to unitarily realize an
  optimal phase covariant transposition map is $d/2$, generalizing the
  result for $d=4$}, for which just a qubit is needed (see Eq.
(\ref{ext-unit-4})). On the other hand, \emph{to span the whole
  optimal maps set one needs a $(d-1)d/2$-dimensional ancilla}.

Finally, notice that realization of multi-phase covariant
transposition generally needs much less resources than realization of
universal transposition: the minimum dimension $d/2$ of the ancilla
space in the phase covariant case has to be compared with the
dimension $d^2$ required in the universal case
(\ref{eq:ancilla-cloning}).

%% file: Chap4.tex
\chapter{The Role of Noise in Quantum Processes}\label{ch:noise}
In the previous Chapters we saw how to optimize transformations over
quantum systems and how to realize them by means of physical
interactions. Of course, processing of quantum systems requires a very
high level of control during all steps of the experiment. On the other
hand, noise---in the sense of uncontrollable interactions of the
system with the sorroundings---is not always and completely avoidable:
the only thing the experimenter can do is to reduce it in order to
reach the desired level of confidence. This can be done by trying to
directly control the environment, e.~g. forcing it into a cavity, or
by engineering states, interactions and measuring apparata robust with
respect to the adopted model of noise.

In the following Sections we will deal with noise on measuring
apparata and on states. While in the first part (review of
Ref.~\cite{cleanpovm}) we will face very general models of
noise---basically, all non-unitary completely positive maps---in the
second part (review of Ref.~\cite{decomap}) we will focus on
decoherence of quantum states, proposing a novel correcting scheme
retrieving classical information that the decoherence process made
leak into the environment and exploiting such information to undo the
noise.
\section{Clean POVM's}
\begin{figure}
\centering
\includegraphics[width=14cm]{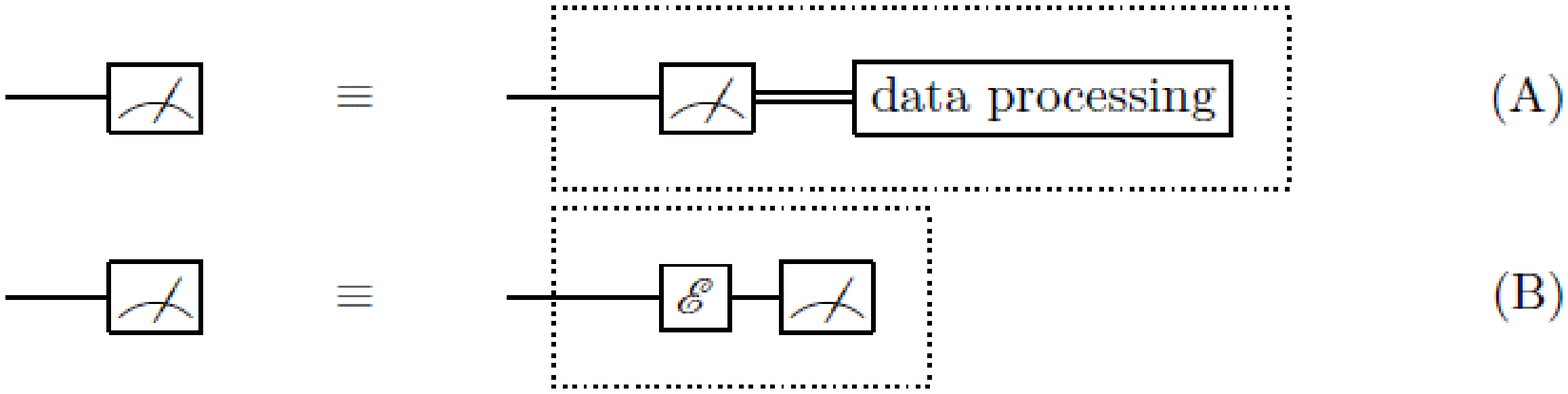}
\caption{There are two ways of processing POVM's: (A) the
  \emph{postprocessing} of the output data and (B) the
  \emph{preprocessing} of the input states by a quantum channel. The
  postprocessing is purely classical, whilst the preprocessing is
  quantum.}\label{fig:noise-in-meas}
\end{figure}

Let us given a general apparatus performing a measurement on quantum
states. We know that the most general way to mathematically model it
is by means of a POVM $\povm P$, see Section~\ref{sec:POVM}. Let us
now think for a while we don't know how the apparatus $\povm P$ works.
It could be noisy at the input gate, that is, quantum states undergo
some uncontrolled transformation $\map E$ before being measured,
and/or noisy at the output, the outcomes being, let's say, shuffled
before being read by the experimenter. The two situations are depicted
in Fig.~\ref{fig:noise-in-meas}.
The question is the following: Do we have any condition on $\povm P$
that allows us to \emph{a priori} affirm that $\povm P$ is ``clean'',
i.~e. without noisy processing at the input and/or the output?

Clearly the point can be viewed from the complementary point of view:
What kind of processings are possible on a given POVM? How does the
apparatus change after such processings are performed?
\subsection{Postprocessing of output data}
The most general postprocessing of a POVM outcomes is a shuffling of
with conditional probability $p(i|j)\ge 0$, corresponding to the
mapping
\begin{equation}
Q_i=\sum_jp(i|j)P_j,
\label{eq:postproc}
\end{equation}
where $\sum_ip(i|j)=1$, $\forall j$. To visualize the shuffling
(\ref{eq:postproc}), it is useful to think to the POVM as a column of
operators and $p(i|j)$ as a column-stochastic matrix\footnote{That is,
  a matrix of positive numbers such that all its columns' entries sum
  up to one.}
\begin{equation}
\begin{pmatrix}
Q_1\\
Q_2\\
\vdots\\
Q_n
\end{pmatrix}=\begin{pmatrix}
p(1|1) & p(1|2) & \cdots & p(1|m)\\
p(2|1) & p(2|2) & \cdots & p(2|m)\\
\vdots & \vdots & \ddots & \vdots\\
p(n|1) & p(n|2) & \cdots & p(n|m)\\
\end{pmatrix}\begin{pmatrix}
P_1\\
P_2\\
\vdots\\
P_m
\end{pmatrix}.
\end{equation}
Notice that postprocessing generally does not require that $\bP$ and
$\bQ$ have the same cardinality. Relevant examples of postprocessing
are:
\begin{enumerate}
\item identification of two outcomes, e.~g. $j$ and $k$ are identified
  with the same outcome $l$, corresponding to
  $p(n|j)=p(n|k)=\delta_{ln}$;
\item permutation $\pi$ of outcomes, corresponding to
  $p(\pi(j)|k)=\delta_{jk}$.
\end{enumerate}

When two POVM's $\bP$ and $\bQ$ are connected by a mapping of the form
\eqref{eq:postproc} for some conditional probability $p(i|j)$ we will
write $\bP\succ_p\bQ$, and say that the POVM $\bP$ is {\em cleaner
  under postprocessing}---{\em postprocessing cleaner}, for
short---than the POVM $\bQ$. It is possible to prove that the relation
$\succ_p $ is a pseudo-ordering, hence an equivalence relation under
postprocessing can be defined as follows
\begin{definition}
  The POVM's $\bP$ and $\bQ$ are {\em postprocessing equivalent}---in
  symbols $\bP\simeq_p \bQ$---if and only if both relations
  $\bP\succ_p\bQ$ and $\bQ\succ_p\bP$ hold.
\end{definition}
We are now in position to define {\em cleanness under postprocessing},
namely
\begin{definition}
  A POVM $\bP$ is {\em postprocessing clean} if and only if for any
  POVM $\bQ$ such that $\bQ\succ_p\bP$, then also $\bP\succ_p\bQ$
  holds, namely $\bP\simeq_p\bQ$.
\end{definition}
The complete characterization of cleanness under postprocessing
(classical) is given by the following theorem (see
Refs.~\cite{cleanpovm,martens})
\begin{theorem}[postprocessing]
  A POVM $\bP$ is postprocessing clean if and only if it is rank-one.
\end{theorem}
This means that if a POVM is rank-one, we are sure that it does not
have a hidden noisy postprocessing at the output. Viceversa, the
Theorem says that it is not possible to obtain the statistics of a
rank-one POVM by classically postprocessing the outcomes of a higher
rank POVM.
\subsection{Preprocessing of input states}
A preprocessing $\map E$ of input states induces naturally a dual
channel $\dual{\map E}$ acting on the POVM itself, as seen in
Subsection~\ref{subsec:quantum-operations}. Hence, we will write
\begin{equation}
\bP\succ\bQ
\end{equation}
and say that the POVM $\bP$ is \emph{preprocessing cleaner} than
$\bQ$, if and only if there exists a channel---i.~e. a
trace-preserving, completely positive map $\map E$---such that
$Q_i=\dual{\map E}(P_i)$, $\forall i$, or, equivalently
$\bQ=\dual{\map E}(\bP)$, for short. It is possible to prove that the
relation $\succ$ is a pseudo-ordering, hence an equivalence relation
under preprocessing can be defined as follows
\begin{definition}
  The POVM's $\bP$ and $\bQ$ are {\em preprocessing equivalent}---in
  symbols $\bP\simeq\bQ$---if and only if both relations $\bP\succ\bQ$
  and $\bQ\succ\bP$ hold.
\end{definition}
We are now in position to define {\em cleanness under preprocessing},
namely
\begin{definition}
  A POVM $\bP$ is {\em preprocessing clean} if and only if for any
  POVM $\bQ$ such that $\bQ\succ\bP$, then also $\bP\succ\bQ$ holds,
  namely $\bP\simeq\bQ$.
\end{definition}

From the above definition, it turns out that a POVM is preprocessing
clean if and only if, whenever a noisy preprocessing acts, its action
on the POVM can be perfectly inverted. Now, a result by Wigner tells
that a channel admits an inverse channel (i.~e. it is physically
invertible\footnote{There exist channels that are invertible in the
  sense that they define a one-to-one correspondence between states,
  but their inverse mappings are not channels. This is the case, for
  example, of the isotropic depolarizing channel $\rho\mapsto
  p\rho+(1-p)I/d$. In Ref.~\cite{cleanpovm} we actually derived, as a
  corollary, that one-to-one channels either are unitary or their
  inverse map is not even positive.}) if and only if such a channel is
actually unitary.  A question arises: Does cleanness property define
an interesting structure in the set of POVM's? Or will we find that
invertible preprocessings are just unitary (i.~e.  trivial)
preprocessings?  Generally, this is not the case, because we want the
action of the noise to be invertible only \emph{on a fixed} POVM, not
on all $\sB(\sH)$.

For qubits, however, preprocessing-equivalence coincides with
unitary-equivalence
\begin{theorem}[qubits] For two-level systems $\bP\simeq\bQ$ if and
  only if there exists a unitary operator $U$ such that $\bP=U^\dag\bQ
  U$.
\end{theorem}

In higher dimensions the counterexample is given implicitly by the
following Theorem regarding effects (two-outcomes POVM's, see
Section~\ref{sec:POVM})
\begin{theorem}[effects]
  Let $\bP=\{P,I-P\}$ and $\bQ=\{Q,I-Q\}$ be two effects. Then
  $\bP\simeq\bQ$ if and only if $\lambda_M(P)=\lambda_M(Q)$ and
  $\lambda_m(P)= \lambda_m(Q)$, where $\lambda_m(O)$ ($\lambda_M(O)$)
  is the minimum (maximum) eigenvalue of $O$.
\end{theorem}
Since necessary and sufficient condition for preprocessing-equivalence
of two effects is that they have the same spectral width, ragardless
of the spectrum itself, it is clear that there exist
preprocessing-equivalent effects which are not unitarily equivalent
(otherwise they should have the same spectrum \emph{as a whole}). It
is also clear that, for dimension $d=2$, the spectrum is completely
determined by the spectral width, whence unitary-equivalence.

Besides effects, the other case in which we have a complete
characterization of preprocessing clean POVM's is the following
\begin{theorem}[observables]\label{th:cleand}
  For number of outcomes $n\leq d$, the set of preprocessing clean
  POVM's coincides with the set of observables.
\end{theorem}
This result is interesting since it provides an operational approach,
alternative to the axiomatic one given by von Neumann, to define what
are the observables in quantum theory. Here, just by introducing the
cleanness pseudo-ordering, we singled out the set of observables, as
the only clean POVM's with numer of outcomes less or equal to the
dimension of the Hilbert space---in this sense, they are the only
clean ``classical'' POVM's.

When the numbers of outcomes gets larger than the dimension of the
Hilbert space, the structure introduced by the preprocessing
pseudo-ordering on the convex set of POVM's becomes more complicated,
and we have just partial results. For example, we can prove that
rank-one POVM's are not only postprocessing clean, but also
preprocessing clean
\begin{theorem}[rank-one]
  Rank-one POVM's are preprocessing clean.
\end{theorem}
Notice that cleanness under preprocessing and extremality are
properties completely unrelated. Consider, e.~g., the following
rank-one POVM
\begin{equation}\label{eq:badpovm}
\frac 12|1\>\<1|,\frac 12|1\>\<1|,|2\>\<2|,\dots,|d\>\<d|.
\end{equation}
The redundantly doubled outcome $|1\>\<1|$ suggests at first sight
that such a POVM cannot be extremal, namely, it cannot be the solution
of any optimization problem. In this sense, such POVM ``is not good''.
However, being rank-one, it is clean under both preprocessing and
postprocessing.
\subsection{Positive maps}
Since now, we introduced two pseudo-orderings on the set of POVM's,
the preprocessing ordering $\succ$, and the postprocessing ordering
$\succ_p$. In this Subsection, we will introduce two additional
relations which can be established among POVM's, namely
\begin{definition}[positive preprocessing] We write $\bP\gg\bQ$ and
  say that $\bP$ is cleaner than $\bQ$ under positive preprocessing,
  if and only if there exists a positive (non necessarily
  \emph{completely} positive) map $\map P$ such that $\bQ=\map
  P(\bP)$.
\end{definition}
\begin{definition}[range-inclusion] We write $\bP\supset_r\bQ$ and say
  that $\bP$ range-includes $\bQ$, if and only if
  $\Rng(\bQ)\subseteq\Rng(\bP)$, where the range of a POVM is defined
  in Definition~\ref{def:POVM-range}.
\end{definition}

We simply have the following hierarchy of relations
\begin{equation}\label{eq:positive-relations}
  \bP\succ\bQ\quad\Longrightarrow\quad\bP\gg\bQ\quad\Longrightarrow\quad\bP\supset_r\bQ.
\end{equation}
The converse is not always true. However, we have some results
providing sufficient conditions for which some of the relations in
Eq.~(\ref{eq:positive-relations}) can be inverted. Proofs are very
technical and can be found in Ref.~\cite{cleanpovm}. Here we just give
the statements.
\begin{theorem}
  Consider two POVM's $\bP$ and $\bQ$ with the same number of
  outcomes. Then the following statements are equivalent:
  \begin{enumerate}
  \item \label{item:1} 
    $\bP \Rip \bQ$
  \item \label{item:2}
    There is a (unique) positive map $\map{E}: \Span(\bP) \to
    \Span(\bQ)$ with $\map{E}(\bP) = \bQ$.
  \end{enumerate}
\end{theorem}
Notice that point~(\ref{item:2}) does not say that $\bP\gg\bQ$, since
the positive map is defined only from $\Span(\bP)$ to $\Span(\bQ)$,
and generally cannot be extended to a positive map on all $\sB(\sH)$.
The following Theorems describe some situations in which it possible
to extend the map $\map E$ to a positive map over all $\sB(\sH)$.
\begin{theorem}
  Consider two POVM's $\bP$ and $\bQ$ with the same number of
  outcomes. Then the following statements are equivalent:
  \begin{enumerate}
  \item \label{item:3a} $\bP \succ \bQ$
  \item \label{item:4} There is an informationally complete POVM $\bM$
    such that $\bP \otimes \bM \Rip \bQ \otimes \bM$.
  \item \label{item:5} $\bP \otimes \bM \Rip \bQ \otimes \bM$ holds
    for all POVM's $\bM$.
  \end{enumerate}
\end{theorem}
\begin{theorem}[abelian POVM]\label{thm:abel-ranges}
  Consider two POVM's $\bP$ and $\bQ$ with the same number of
  outcomes. Let $\bQ$ be abelian, namely $Q_iQ_j=Q_jQ_i$ for all
  $i,j$. Then $\bP \Rip \bQ$ $\Longrightarrow$ $\bP \succ \bQ$, and
  Eq.~(\ref{eq:positive-relations}) becomes a chain of equivalences.
\end{theorem}
\section{Inverting decoherence}
We will now focus our attention on a particularly nasty preprocessing
of input states, namely on decoherence. Decoherence is universally
considered, on one side, as the major practical limitation for
communication and processing of quantum information. On the other
side, decoherence yields the key concept to explain the transition
from quantum to classical world \cite{decoherence} due to the
uncontrolled and unavoidable interactions with the environment. Great
effort in the literature has been devoted to combat the effect of
decoherence by engineering robust encoding-decoding schemes. Some
authors have recently addressed a different approach to undo quantum
noises by extracting classical information from the environment
\cite{GW} and exploiting it as an additional amount of side
information useful to improve quantum communication performances
\cite{capacity}.

The recovery of quantum coherence from the environment is often a
difficult task, e. g.  when the environment is ``too big'' to be
controlled, as for spontaneous emission of radiation. By regaining
control on the environment the recovery can sometimes be actually
accomplished, for example by keeping the emitted radiation inside a
cavity. However, in some cases, the full recovery of quantum coherence
becomes impossible even in principle, namely even when one has
complete access to the environment. This naturally leads us to pose
the following question: in which physical situations is possible to
perfectly recover quantum coherence by monitoring the environment?

\subsection{Convex structure of decoherence maps}

A completely decohering evolution asymptotically cancels any quantum
superposition when reaching the stationary state, making any state
diagonal in some fixed orthonormal basis---the basis depending on the
particular system-environment interaction.  In the Heisenberg picture
we say that such a completely decohering evolution asymptotically maps
the whole algebra of quantum observables into a ``maximal classical
algebra'', that is a maximal set of commuting---namely jointly
measureable---observables. Let's denote by $\alg A_q$ the ``quantum
algebra'' of all bounded operators $\sB(\sH)$ on the finite
dimensional Hilbert space $\sH$, and by $\alg A_c$ the ``classical
algebra'', namely any maximal Abelian subalgebra $\alg A_c\subset\alg
A_q$.  Clearly, all operators in $\alg A_c$ can be jointly
diagonalized on a common orthonormal basis, which in the following
will be denoted as $\Base =\{|k\>|k=1, \dots ,d\}$. Then, the
classical algebra $\alg A_c$ is also the linear span of the
one-dimensional projectors $|k\>\<k|$, whence $\alg A_c$ is a
$d$-dimensional vector space. According to the above general
framework, we call \emph{(complete) decoherence map} a completely
positive identity-preserving (i.~e. trace-preserving in the
Schr\"odinger picture, see Subsection~\ref{subsec:quantum-operations})
map $\dual{\map E}$ which asymptotically maps any observable $O \in
\alg A_q$ to a corresponding ``classical observable'' in $\alg A_c$,
namely such that the limit $\lim_{n\to \infty} (\dual{\map E})^n (O)$
exists and belongs to the classical algebra $\alg A_c$ for any $O \in
\alg A_q$. Here we denote with $(\dual{\map E})^n$ the $n$-th
iteration of the map $\map E$, implicitly assuming markovian
evolution.

It is easy to see that the set of decoherence maps is convex. The
following Theorem shows that such maps enjoy a remarkably simple form:
\begin{theorem}[Schur form]\label{GeneralForm}
  A map $\dual{\map E}$ preserves all elements of the maximal
  classical algebra $\alg A_c$ if and only if it has the form
\begin{equation}\label{SchurMap}
  \dual{\map E}(O)= \xi \circ O,
\end{equation}
$A \circ B$ denoting the Schur product of operators $A$ and $B$, i.~e.
$A \circ B \equiv \sum_{k,l=1}^d A_{kl} B_{kl} |k\>\<l|$, $\{A_{kl}\}$
and $\{B_{kl}\}$ being the matrix elements of $A$ and $B$ in the basis
$\Base$, and $\xi_{kl}$ being a correlation matrix, i.e. a positive
semidefinite matrix with $\xi_{kk}=1$ for all $k=1, \dots ,d$.
\end{theorem}
Theorem~\ref{GeneralForm} states a linear correspondence between maps
preserving $\alg A_c$ and correlation matrices, whence the two sets
share the same convex structure. Then the map is extremal if and only
if its correlation matrix is extremal.

Since now we dealt with the dual map $\dual{\map E}$ on bounded
operators. The action of a decoherence map on quantum states is given
in Schr\"odinger picture by
\begin{equation}\label{SchurFormSchro}
  \map E(\rho)= \xi^T \circ \rho,
\end{equation}
where $T$ denotes transposition with respect to the basis $\Base$
(also $\xi^T$ is a correlation matrix, hence in the following, we will
drop the symbol $T$ at the exponent). As a consequence, one has
exponential decay of the off-diagonal elements of $\rho$, since
$\left|[\map E^n (\rho)]_{kl}\right|=|\xi_{kl}|^n\cdot |\rho_{kl}|$.
In other words, any initial state $\rho$ decays exponentially towards
the completely decohered state
\begin{equation}
  \rho_\infty\equiv\sum_k\rho_{kk}|k\>\<k|.
\end{equation}

In Ref.~\cite{decomap}, it is proved the following
\begin{lemma}\label{l:extr}
  A map $\map E$ is an extremal decoherence map if and only if it is
  extremal in the set of all maps.
\end{lemma}
As a consequence of Lemma \ref{l:extr}, the convex structure of
decoherence maps can be obtained by application of the well known Choi
Theorem \cite{choi}, which states that the canonical Kraus
operators\footnote{For the definition of canonical Kraus
  decomposition, see Subsection~\ref{subsec:quantum-operations}.}
$\{E_i\}$, $1\leq i \leq r$, of every extremal map are such that their
products $\{E_i^{\dag}E_j\}$, $1\leq i,j\leq r$, are linearly
independent. A relevant consequence of this characterization is the
following
\begin{theorem}
  If $\map E$ is an extremal decoherence map, then $r \leq \sqrt d$.
  For qubits and qutrits any decoherence map is then random-unitary.
\end{theorem}
This means that for qubits and qutrits extremal decoherence maps are
unitary maps, since they admits a Kraus representation containing only
one operator. Hence, for qubits and qutrits, every decoherence map can
be written as
\begin{equation}\label{RandomUnitary}
  \map E(\rho)=\sum_ip_iU_i\rho U_i^\dag,
\end{equation}
for some commuting unitary operators $U_i\in\alg A_c$ and probability
distribution $p_i$.
\subsection{Correcting decoherence by measuring the environment}
In Ref.~\cite{GW} it is shown that the only channels that can be
perfectly inverted by monitoring the environment are the
random-unitary ones. Therefore, it follows that one can perfectly
correct any decoherence map for qubits and qutrits by monitoring the
environment.  The correction is achieved by retrieving the index $i$
in Eq.~(\ref{RandomUnitary}) via a measurement on the environment, and
then by applying the inverse of the unitary transformation $U_i$ on
the system. Therefore, the random-unitary map simply leaks $H(p_i)$
bits of \emph{classical information} into the environment ($H$
denoting the Shannon entropy), and the effects of decoherence can be
completely eliminated by recovering such classical information,
without any prior knowledge about the input state. The fact that
decoherence maps are necessarily random-unitary is true only for
qubits and qutrits. A counterexample in dimension $d=4$ can be found
in Ref.~\cite{decomap}. Such extremal decoherence maps with $r\geq2$
represent a process which is fundamentally different from the random
unitary one, corresponding to a {\em leak of quantum information} from
the system to the environment, information that cannot be perfectly
recovered from the environment \cite{GW}.

Now we address the problem of estimating the amount of classical
information needed in order to invert a random-unitary decoherence
map. If the environment is initially in a pure state, say $|0\>_e$, a
useful quantity to deal with is the so-called entropy
exchange \cite{Schumacher} $\Sex$ defined as
\begin{equation}\label{DefSex}
  \Sex(\rho)=S(\sigma_{e}^\rho),
\end{equation}
where $\sigma_{e}^\rho$ is the reduced environment state after the
interaction with the system in the state $\rho$, and
\mbox{$S(\rho)=-\Tr[\rho\log\rho]$} is the von Neumann entropy. In
the case of initially pure environment, the entropy exchange depends
only on the map $\map E$ and on the input state of the system $\rho$,
regardless of the particular system-environment interaction chosen to
model $\map E$. It quantifies the information flow from the system to
the environment and, for all input states $\rho$, one has the
bound \cite{Schumacher} $|S(\map E(\rho))-S(\rho)|\leq \Sex(\rho)$,
namely the entropy exchange $\Sex$ bounds the entropy production at
each step of the decoherence process.

In order to explicitly evaluate the entropy exchange for a decoherence
process, we can then exploit a particular model interaction between
system and environment. This can be done noticing that it is always
possible to write $\xi_{kl}=\<e_l|e_k\>$ for a suitable set of
normalized vectors $\{|e_k\>\}$.  Then, the map $\map
E(\rho)=\xi\circ\rho$ can be realized as \mbox{$\map
  E(\rho)=\Tr_e[U(\rho\otimes |0\>\<0|_e)U^\dag]$}, where the unitary
interaction $U$ gives the transformation
\begin{equation}\label{Unitary}
U|k\>\otimes|0\>_e=|k\>\otimes|e_k\>,
\end{equation}
whence the final reduced state of the environment is
$\sigma_{e}^\rho=\sum_k\rho_{kk}|e_k\>\<e_k|$. Then, in order to
evaluate $\Sex$ for a decoherence map $\map E(\rho)=\xi\circ\rho$, it
is possible to bypass the evaluation of the states $|e_i\>$ of the
environment, using the formula
\begin{equation}\label{Sex_and_xi}
  \Sex(\rho)=S(\sqrt{\rho_\infty}\xi\sqrt{\rho_\infty}),
\end{equation}
which follows immediately from the fact that
$\sqrt{\rho_\infty}\xi\sqrt{\rho_\infty}$, and $\sigma_{e}^\rho$ are
both reduced states of the same bipartite pure state
$\sum_i\sqrt{\rho_{ii}}|i\>|e_i\>$.

Notice that the unitary interaction $U$ in Eq.~(\ref{Unitary})
generalizes the usual form considered for quantum
measurements \cite{vonneumann}, with the quantum system interacting
with a pointer, which is left in one of the (nonorthogonal) states
$\{|e_k\>\}$. The more the pointer states are ``classical''---i.~e.
distinguishable---the larger is the entropy exchange, whence the
faster is the decoherence process. In the limit of orthogonal states,
decoherence is istantaneous, i.~e. $\map E(\rho)=\rho_\infty$.

When a map can be inverted by monitoring the environment---i.~e. in
the random-unitary case---the entropy exchange $\Sex(I/d)$ provides a
lower bound to the amount of classical information that must be
collected from the environment in order to perform the correction
scheme of Ref.~\cite{GW}. In fact, assuming a random-unitary
decomposition (\ref{RandomUnitary}) and using the formula
\cite{Schumacher} $\Sex (\rho)=S\left( \sum_{i,j} \sqrt{p_i p_j}
  \Tr[U_i \rho U_j^{\dag}] |i\>\<j| \right)$, we obtain
\begin{equation}\label{SexAndEntropy}
\Sex(I/d) \leq H(p_i).
\end{equation} 
The inequality comes from the fact that the diagonal entries of a
density matrix are always majorized by its eigenvalues \cite{bhatia},
and it becomes equality if and only if
$\Tr[U_iU_j^{\dag}]/d=\delta_{ij}$, i.~e. the map admits a
random-unitary decomposition with \emph{orthogonal} unitary operators.
Moreover, from Eq. (\ref{Sex_and_xi}) we have $\Sex
(I/d)=S(\xi/d)$.

In Ref.~\cite{decomap}, it is proved that, for qubits, $S(\xi/2)$
quantifies exactly the minimum amount of classical information which
must be extracted from the environment, while, for dimension $d>2$,
the bound in Eq.~(\ref{SexAndEntropy}) is generally strict and a
counterexample is given for dimension $d=3$. Notice that the same
decoherence map may be obtainable by different random-unitary
decompositions with different probability distributions $\{p_i\}$,
corresponding to different values of the information $H(p_i)$.
However, for qubits it is always possible to perform a suitable
measurement on the environment and to invert the decoherence map
retrieving the \emph{minimal} amount of information from the
environment, namely $S(\xi/2)$. For example, consider the so-called
\emph{random phase-kick model} \cite{nielsen} for decoherence of
qubits
\begin{equation}\label{badrandomphase}
  \map E(\rho)=\frac{1}{\sqrt{4\pi\lambda}}\int_{-\infty}^{+\infty}e^{i\theta\sigma_z/2}\rho e^{-i\theta\sigma_z/2}e^{-\theta^2/4\lambda}\d\theta,
\end{equation}
which can be rewritten as
\begin{equation}\label{goodphasekick}
\map E(\rho)=\xi\circ\rho,\qquad
\xi=\begin{pmatrix}
1 & e^{-\lambda}\\
e^{-\lambda} & 1
\end{pmatrix}.
\end{equation}
From Eq.~(\ref{badrandomphase}), one could infer that the amount $S$
of classical information that must be extracted from the environment
is equal to the differential entropy of the gaussian probability
density according to which the system is random phase-kicked, namely
(see Ref.~\cite{cover-thomas})
\begin{equation}\label{eq:wronginfo}
  S\left(\frac{1}{\sqrt{4\pi\lambda}}e^{-\theta^2/4\lambda}\right)=\frac{1}{2}\log4\pi e\lambda,
\end{equation}
growing logarithmically with $\lambda$. This is actually not correct,
since the \emph{minimum} amount of classical information needed is
\begin{equation}\label{eq:entropyrate}
  S=-p\log_2p-(1-p)\log_2(1-p),\qquad p=\frac{1-e^{-\lambda}}{2}.
\end{equation}
In fact, $\xi/2$ in Eq.~(\ref{goodphasekick}) can be simply
diagonalized and has eigenvalues
$\left\{\frac{1-e^{-\lambda}}{2},\frac{1+e^{-\lambda}}{2}\right\}$.
\begin{figure}
\centering
\includegraphics[width=14cm]{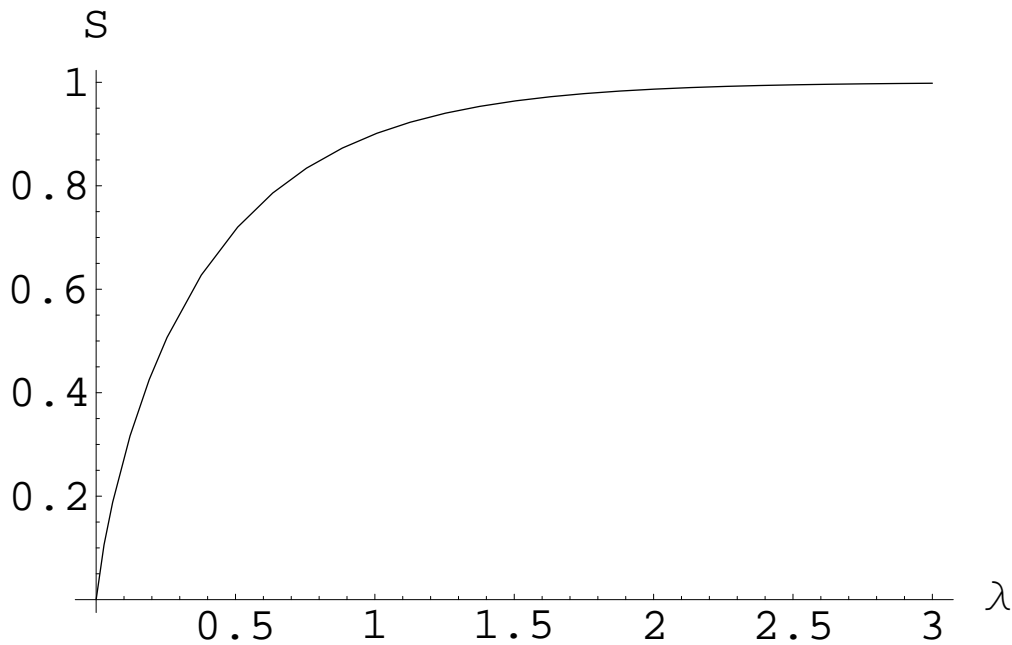}
\caption{The amount of classical information $S$, expressed in bits,
  leaking into the environment at every application of the random
  phase-kick model of decoherence for qubits, as function of the
  parameter $\lambda$, see Eqs.~(\ref{badrandomphase})
  and~(\ref{goodphasekick}). $S$ tends to the limit value of 1 bit,
  since \emph{every} qubit decoherence map can be written as a
  random-unitary process involving only \emph{two} unitaries (see the
  footnote in the previous page).}\label{fig:entropyrate}
\end{figure}
In Figure~\ref{fig:entropyrate} there is the plot of the amount of
classical information $S$ in Eq.~(\ref{eq:entropyrate}) as a function
of the parameter $\lambda$ modelling the decoherence rate in
Eqs.~(\ref{badrandomphase}) and~(\ref{goodphasekick}). The curve tends
to the finite limit of one bit, contrarily to what happens in
Eq.~(\ref{eq:wronginfo}).

%% file: conclusions.tex
\chapter*{List of Publications}
\addcontentsline{toc}{chapter}{\bf List of Publications}
\begin{itemize}
\item F~Buscemi, G~M~D'Ariano, C~Macchiavello, and P~Perinotti,\\
  \emph{Optimal superbroadcasting maps of mixed qubit states},\\
  in preparation
\item F~Buscemi, G~M~D'Ariano, M~Keyl, P~Perinotti, and R~F~Werner,\\
  \emph{Clean positive operator valued measures},\\
  J. Math. Phys. {\bf 46}, 082109 (2005)
\item F~Buscemi, G~Chiribella, and G~M~D'Ariano,\\
  \emph{Inverting quantum decoherence by classical feedback
    from the environment},\\
  Phys. Rev. Lett. {\bf 95}, 090501 (2005)
\item F~Buscemi, G~M~D'Ariano, and C~Macchiavello,\\
  \emph{Optimal Time-Reversal of Multi-phase Equatorial States},\\
  pre-print on \texttt{quant-ph/0504016}
\item F~Buscemi, G~M~D'Ariano, and C~Macchiavello,\\
  \emph{Economical Phase-Covariant Cloning of Qudits},\\
  Phys. Rev. A {\bf 71}, 042327 (2005)
\item F~Buscemi, G~M~D'Ariano, and P~Perinotti,\\
  \mbox{\emph{There exist non orthogonal quantum measurements that are
      perfectly repeatable},}\\
  Phys. Rev. Lett. {\bf 92}, 070403 (2004)
\item F~Buscemi, G~M~D'Ariano, and M~F~Sacchi,\\
  \emph{Physical realizations of quantum operations},\\
  Phys. Rev. A {\bf 68}, 042113 (2003)
\item F~Buscemi, G~M~D'Ariano, P~Perinotti, and M~F~Sacchi,\\
  \emph{Optimal realization of the transposition maps},\\
  Phys. Lett. A {\bf 314}, 374 (2003)
\item F~Buscemi, G~M~D'Ariano, and M~F~Sacchi,\\
  \emph{Unitary realizations of the ideal phase measurement},\\
  Phys. Lett. A {\bf 312}, 315 (2003)
\end{itemize}